\RequirePackage{fix-cm}
\documentclass[twocolumn]{svjour3}          
\smartqed  
\usepackage{graphicx}
\usepackage{times}
\usepackage{epsfig}
\usepackage{epstopdf}
\usepackage{amsmath}
\usepackage{amsfonts}
\usepackage{amssymb}
\usepackage{subfigure}
\usepackage{caption}
\usepackage{tabularx}
\usepackage[table]{xcolor}
 \usepackage[normalem]{ulem}
\usepackage{algorithm}
\usepackage[noend]{algpseudocode}

\usepackage[section]{placeins}

\usepackage{graphicx}
\usepackage{color}
\usepackage{overpic}
\usepackage{color}
\usepackage{algorithm}
\usepackage{algpseudocode}
\usepackage{caption}
\usepackage{newfloat}
\usepackage{placeins}
\usepackage{setspace}
\usepackage{array}
\usepackage{enumitem}
\usepackage{appendix}
\usepackage{tabularx}
\usepackage{xspace}

\DeclareMathOperator*{\trace}{tr}

\DeclareFloatingEnvironment[fileext=lod]{diagram}
\DeclareRobustCommand\onedot{\futurelet\@let@token\@onedot}
\def\@onedot{\ifx\@let@token.\else.\null\fi\xspace}

\newcommand{\norm}[1]{\left\lVert#1\right\rVert}
\newcommand{\abs}[1]{\lvert#1\rvert}
\newcommand*{\defeq}
\makeatletter
\def\BState{\State\hskip-\ALG@thistlm}
\makeatother
\definecolor{myGreen}{HTML}{33FF00}
\definecolor{myRed}{HTML}{FF3030}
\definecolor{myGrey}{HTML}{AA5555}
\definecolor{myWhite}{HTML}{FFFFFF}
\definecolor{maroon}{cmyk}{0,0.87,0.68,0.32}
\definecolor{petr}{HTML}{5555FF}
\definecolor{josef}{HTML}{FF3030}

\begin{document}

\title{Fast Blended Transformations for Partial Shape Registration}


\author{Alon Shtern\textsuperscript{*} \and Matan Sela\textsuperscript{*} \and Ron Kimmel\\
Technion - Israel Institute of Technology\\
{\tt\small ashtern@campus.technion.ac.il}\\ {\tt\small matansel@cs.technion.ac.il}\\{\tt\small ron@cs.technion.ac.il}}


\date{Received: date / Accepted: date}

\maketitle

  \let\thefootnote\relax\footnote{*Equal contribution}
\begin{figure}[t]
	\begin{center}
	\begin{overpic}[width=1\columnwidth]{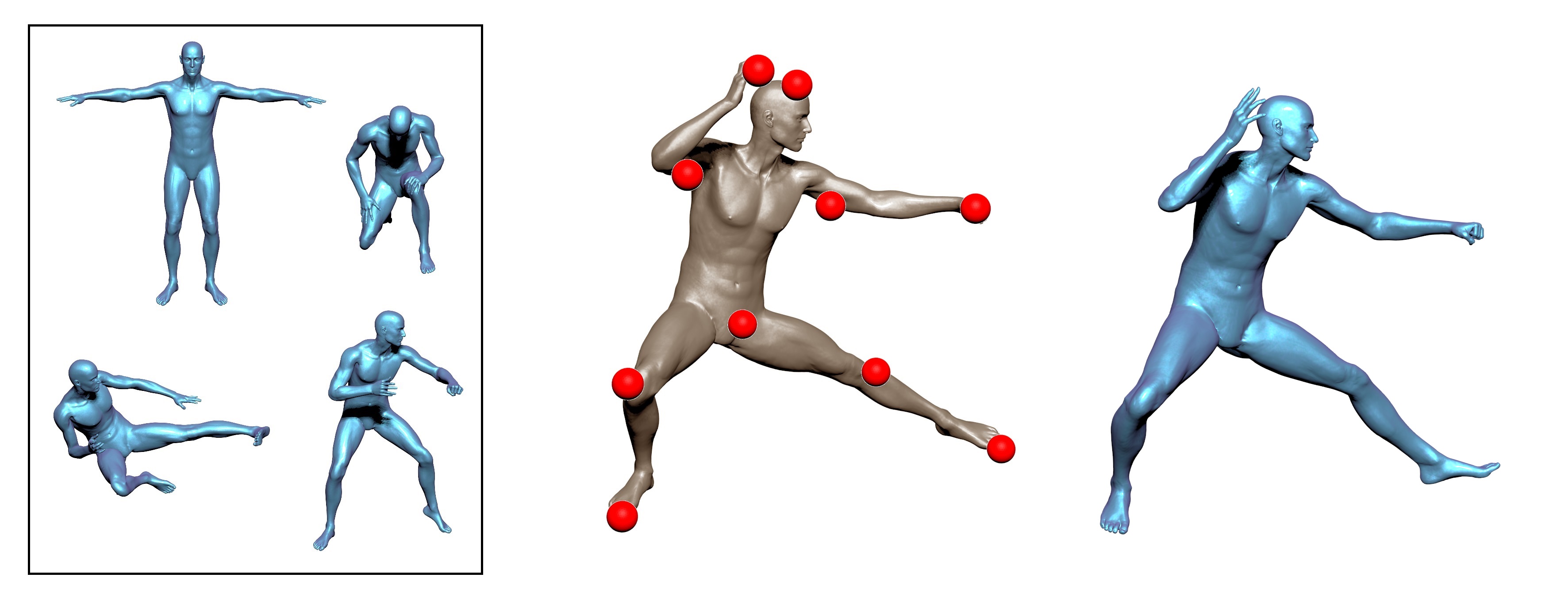}
	\end{overpic}
	\end{center}
	\vspace{-2em}
	\caption{\small A shape obtained by blended transformation using ten positional
               constraints and four reference shapes.}
\end{figure}

\begin{abstract}
  Automatic estimation of skinning transformations is a popular way to deform
   a single reference shape into a new pose by providing a small number of control parameters.
  We generalize this approach by efficiently enabling the use
   of multiple exemplar shapes.
  Using a small set of representative natural poses,
   we propose to express an unseen appearance by a low-dimensional
   linear subspace, specified by a redundant dictionary of weighted vertex positions.
  Minimizing a nonlinear functional that regulates the example manifold,
   the suggested approach supports local-rigid deformations of articulated objects,
   as well as nearly isometric embeddings of smooth shapes.
  A real-time non-rigid deformation system is demonstrated, and a shape
   completion and partial registration framework is introduced.
  These applications can recover a target pose and implicit inverse kinematics
   from a small number of examples and just a few vertex positions.
  The result reconstruction is more accurate compared to state-of-the-art
   reduced deformable models.

\end{abstract}

\section{Introduction}
\label{intro}
The construction of an efficient automatic procedure that deforms one shape
into another in a natural manner is a fundamental and well-studied challenge
in computer graphics.
Professional animators design deformable models for manually editing
facial expressions, controlling postures and muscles of shapes, and creating
sequences of gestures and motions of animated objects.
Such models also play a key role in the field of shape analysis.
For example, elastic surface registration techniques try to iteratively
warp given shapes so as to establish an optimal alignment between them.

A major challenge in automatic shape deformation is preserving the expressiveness
of the model while reducing its complexity.
This can be accomplished by exploiting the potential redundancy in natural motions.
For instance, in non-rigid articulated objects as hands, the bending of a single
finger mainly influences the movement of nearby skin.
The stiffness of the limbs restricts them to move freely and therefore the
deformation of a shape as a whole can often be well approximated as a blend
of a small number of affine transformations.
One such skeletal deformation technique,
the {\it Linear Blend Skinning} (LBS) \cite{magnenat1988joint},
has been widely adopted by the gaming and the film industries
due to its simplicity and efficiency.

More recently, Jacobson et al. \cite{jacobson2012fast} suggested to deform a
single shape by looking for transformations that minimize the nonlinear
{\it As-Rigid-As-Possible} (ARAP) energy  \cite{alexa2000rigid,sorkine2007rigid}.
This energy penalizes deviations from rigidity of the underlying structural skeleton.
The optimization process alternates between finding the minimal affine
transformations and projecting them onto the group of rigid ones.
The algorithm converges after a few iterations and provides realistic
deformations with a low computational effort.
The method was designed for modifying a single reference shape.
As such, it does not effectively incorporate the nature of plausible non-rigid
deformations that can be well captured by a few examples.
Therefore, this method requires a manually tailored pre-computation of biharmonically
smooth blending functions, and relies on an initial pose of the shape that is
usually selected as the previous frame in the motion sequence.

In many situations, while analyzing or synthesizing shapes, neither manual input nor
the temporal state of the shape at the previous frame is available.
In these circumstances, obtaining a natural initial pose for the nonlinear
optimization procedure becomes a challenge.
Nevertheless, in many of these events, static poses of the \emph{same} shape might be
available, such as in \cite{bogo2014faust} where several human bodies in various
postures were captured and reconstructed using range scanners.
In this paper, we present an efficient generalization 
of the LBS model for the case where multiple exemplar shapes are available.
To that end, the proposed framework uses the reference shapes to infer an expressive
yet low dimensional model, which is computationally efficient and produces
natural looking poses.
The proposed method constructs a dictionary that contains prototype signal-atoms
of weighted vertex coordinates, that effectively span the space of deformations
represented by the exemplar shapes.
We refer to \cite{elad2010sparse}, for applications of overcomplete dictionaries
for sparse and redundant data representations in other domains.

{\color{black}
The proposed algorithm is mainly motivated by the nonrigid 3D partial registration problem.
This problem is considered a key challenge in the field of shape analysis.
One of the most efficient approaches to solve this challenge is using
deformation-driven correspondences \cite{zhang2008deformation}.
A good deformation method for this purpose should efficiently produce
plausible deformations that fits some known constraints.
In our setting we use several example shapes and a few known vertex positions.
Although some example-based methods produce excellent deformations, in this context
of partial registration, they usually carry three major drawbacks.
First, most of these methods have high complexity.
Second, they depend strongly on a good initial shape alignment.
Third, they require many examples for constructing a model which plausibly captures
various poses.
The proposed method tries to overcome these difficulties by using a redundant dictionary
that spans a linear deformation subspace.
The advantages of using a linear subspace are evident.
Acceleration in this case is well established using the ARAP energy functional.
Additionally,  well known regularization techniques, such as $L_1$ and $L_2$ penalty terms,
can easily be deployed in conjunction with the linear model to find a robust sparse
representation for the initial shape alignment.
Moreover, the simplicity and flexibility of using the linear representation enables
the proposed algorithm to refine this initial shape deformation by gradually
expanding the deformation space while simultaneously introducing more
accurate model constraints.}

The key contributions of the proposed approach include the following features.
\begin{itemize}
\item Given a few reference shapes, we construct a redundant, yet compact,
dictionary of weighted positional-atoms that spans a rich space of deformations.
A new deformation is represented as a linear combination of these atom-signals.
\item Stable transformations are established by using sparse modeling over a
limited  subspace of deformations.
The suggested framework ensures the use of only a few dictionary atoms relating
a few given poses to a target one.
\item The As-Rigid-As-Possible energy is reformulated to support multiple
 reference shapes and automatic global scale detection.
\item Smooth deformations are realized by an additional biharmonic energy term
that is computationally efficient to minimize when the skinning weights
are set to be the eigenfunctions of the Laplace-Beltrami operator.
\end{itemize}

To demonstrate the fast blended transformations approach, animation sequences were
generated given just a few reference shapes and a handful of point constraints that
define each target frame.
Quantitative evaluation indicates that the advantages of the proposed approach are
fully realized when plugged into a shape completion and registration application
that achieves low correspondence errors and deformation distortions.

\section{Related efforts}
Example-based deformation techniques attempt to establish a compact representation
 of shape deformations while trying to satisfy desirable properties.
Forming these representations generally requires the processing of sets of poses,
 expressions, or identities of the same class of shapes.
To fulfill this task, various methods have been proposed.
Roughly speaking, they all share the following taxonomy.

\textbf{Displacement field interpolation.}
This technique computes the pointwise difference between each example
 shape and a reference one at a resting pose, see for example
  \cite{lewis2000pose,sloan2001shape,kry2002eigenskin}.
More recent methods include statistical  \cite{feng2008real}
 and rotational regressions \cite{wang2007real}.

\textbf{Deformation gradient.}
These methods interpolate the example poses using the gradient fields of the
 coordinate functions, and construct the deformed surface by solving a Poisson equation.
In \cite{xu2006poisson,sumner2005mesh} the deformation is estimated
 for each triangle of the given mesh.
Example based deformation gradients and its variants, like the Green strain tensor, are also used for static or dynamic simulation of  elastic materials  \cite{martin2011example,koyama2012real,bouaziz2014projective,schumacher2012efficient,zhang2015real}.
For lowering the computational cost, Der et al. \cite{der2006inverse} proposed
 to cluster triangles that are subject to a similar rigid rotation with respect to a single reference shape.
It allowed reformulating the problem in terms of transformations of a representative
 proxy point for each group of vertices.

\textbf{Edge lengths and dihedral angles interpolation.}
Inspired by discrete shells \cite{grinspun2003discrete}, local properties were used
 for  mesh interpolation \cite{winkler2010multi}, that naturally fits with the
 discrete shell energy for combined physics-based and example-driven mesh deformations
 \cite{frohlich2011example}.

\textbf{Transformation blending.}
This approach describes the deformation by a set of affine transformations that
 are blended together to represent the deformed shape.
In this case, the example shapes are used to find the skinning weights as well
 as the transformations by using non-linear optimization algorithms
 \cite{james2005skinning,kavan2010fast,le2014robust,levi2015smooth}.

 %

\textbf{Linear subspace.}
Similar in its spirit to the proposed approach is 
  Tycowicz et al. \cite{tycowicz2015interpolation}.
Their method computes an example-based reduced linear model for representing the
 high dimensional shape space using deformation energy derivatives and Krylov sequences.
However, their framework and reduced linear subspace
 are specifically designed and restricted to the nonlinear shape interpolation problem.

The fast blended transformations method is affiliated with the class of transformation blending
  inspired by \cite{jacobson2012fast,wang2015linear}.
The deformation is performed by minimizing a nonlinear energy functional
 over the linear subspace of skinning transformations.
Unlike previous efforts, we suggest to simultaneously blend affine transformations
 of several given poses of the same subject.
%
The proposed framework allows us to learn the example manifold without
 estimating the explicit connections between the reference shapes.
With these reference shapes, we construct an overcomplete dictionary that spans
 the space of allowed deformations up to a small tolerance.
The nonlinear energy functional guides the transformations to achieve a
 physically-plausible deformation.
Projecting a small set of constraints to the examples manifold,
 which is assumed to be of low dimensions, we obtain an efficient and
 accurate blending procedure for real time animation and for the partial shape registration task.

 \section{Notations and problem formulation}
 \subsection {Linear blend skinning}
 Here, we follow the blend skinning model
  as described by Jacobson et al. in \cite{jacobson2012fast}.
 Let ${\bf v}_1, \dots ,{\bf v}_n \in \mathbb{R}^{d}$ ($d = 3$)
 be the vertex positions of the input reference mesh $\mathcal{M}$ with
  $f$ triangles and $n$ vertices.
  Denote the deformed vertex positions of a new target
   mesh $\tilde{\mathcal{M}}$ by $\tilde{\bf v}_1, \dots, \tilde{\bf v}_n \in \mathbb{R}^{d}$.
   The target vertex positions relate to the given reference vertices through $m$ affine transformation matrices ${\bf M}_j \in  \mathbb{R}^{d \times (d+1)}$, $j=\{1, \dots, m\}$ and real-valued skinning weight functions $w_j$,
 that measure the influence of each affine transformation on each point of the shape.
For a discrete mesh, we denote $w_j({\bf v}_i)$ by $w_{j,i}$, and readily have
\begin{align}
\label{eq:lbs}
	{\bf \tilde{v}}_i =& \sum_{j=1}^m w_{j,i} {\bf M}_j \begin{pmatrix}{\bf v}_i\\ 1\end{pmatrix}.
\end{align}
\hbox{Equation (\ref{eq:lbs})} can be rewritten in a matrix form as
\begin{align*}
\tilde{\bf V} =& {\bf {D}}_{\textsc{\tiny LBS}}{\bf {T}}_{\textsc{\tiny LBS}},
\end{align*}
where $\tilde{\bf V} \in \mathbb{R}^{n \times d}$ is the matrix whose rows are the
positions of the target vertices,
and the matrices ${\bf {T}}_{\textsc{\tiny LBS}}
\in  \mathbb{R}^{ (d+1)m \times d}$ and
${\bf {D}}_{\textsc{\tiny LBS}} \in \mathbb{R}^{n \times (d+1)m}$ are created by stacking the skinning parameters in
the following fashion

\begin{align*}
{\bf {D}}_{\textsc{\tiny LBS}} = \begin{pmatrix} w_{1,1}
   \begin{pmatrix}{\bf v}_1^{\mathrm{T}}, 1
   \end{pmatrix} & \dots  &w_{m,1}  \begin{pmatrix}{\bf v}_1^{\mathrm{T}}, 1 \end{pmatrix}
    \\ \vdots & \ddots & \vdots \\ w_{1,n}
   \begin{pmatrix}{\bf v}_n^{\mathrm{T}}, 1 \end{pmatrix}
      & \dots &w_{m,n}
      \begin{pmatrix}{\bf v}_n^{\mathrm{T}}, 1 \end{pmatrix}
      \end{pmatrix},
\end{align*}
\begin{align*}
    {\bf {T}}_{\textsc{\tiny LBS}} = \begin{pmatrix} {\bf M}_1 & \hdots & {\bf M}_m\end{pmatrix}^{\mathrm{T}}.
\end{align*}

\subsection {Fast automatic skinning transformations}
The most general form of representing the position of a new target vertex by a
 linear transformation of some dictionary (such as the linear blend skinning formulation) can be expressed by
\begin{align*}
\tilde{\bf V} =& {\bf D}{\bf T},
\end{align*}
where ${\bf D} \in \mathbb{R}^{n \times b}$ is a dictionary of size $b$ (in case of standard linear blend skinning $b=(d+1)m$), and ${\bf T} \in \mathbb{R}^{b \times d}$ is a matrix of unknown coefficients that represents the vertex positions in terms of the dictionary.

Jacobson et al. \cite{jacobson2012fast} introduced a method for automatically finding the skinning transformations
 ${\bf T}$ by minimizing the ARAP energy \cite{sorkine2007rigid,liu2008local,chao2010simple} between the reference
 shape $\mathcal{M}$ and the target one $\tilde{\mathcal{M}}$.
Let ${\bf R}_1, {\bf R}_2, \dots ,{\bf R}_r \in \text{SO}(d)$ and  $\mathcal{E}_1, \mathcal{E}_2, \dots, \mathcal{E}_r$ be $r$ local rotations and
their corresponding edge sets, respectively. The ARAP energy, which measures local deviation from
rigidity, can be expressed as
\begin{align*}
E({\bf V},\tilde{\bf V}) =& \cfrac{1}{2}\sum\limits_{k=1}^r\sum\limits_{(i,j)
  \in\mathcal{E}_k} c_{ijk}\|(\tilde{\bf v}_i-\tilde{\bf v}_j)
   -{\bf R}_k({\bf v}_i-{\bf v}_j)\|^2,
\end{align*}
where $c_{ijk} \in \mathbb{R}$ are the cotangent weighting coefficients \cite{pinkall93}.
As indicated  in \cite{jacobson2012fast}, it is unnecessary to estimate
 the local rotation for each edge separately since vertices undergoing similar
 deformations can be clustered together into a small number of rotation clusters.

The ARAP energy can be expressed in a simple matrix form.
Denote ${\bf A}_k \in \mathbb{R}^{n\times \abs{\mathcal{E}_k}}$
 as the directed incidence matrix corresponding to edges $\mathcal{E}_k$,
  and let ${\bf C}_k \in \mathcal{R}^{\abs{\mathcal{E}_k} \times \abs{\mathcal{E}_k}}$
  be a diagonal matrix with weights $c_{ijk}$.
Then, the ARAP energy can be written in matrix form as
\begin{align*}
	2E({\bf V},\tilde{\bf V}) =& \trace(\tilde{\bf V}^{\mathrm{T}} {\bf L} \tilde{\bf V})
    - 2\trace({\bf R} {\bf K} \tilde{\bf V}) +  \trace({\bf V}^{\mathrm{T}} {\bf L} {\bf V}),
\end{align*}
 where ${\bf R} = ({\bf R}_1,\dots,{\bf R}_r)$, ${\bf K} \in {\mathbb R}^{dr \times n}$
 stacks the matrices ${\bf V}^{\mathrm{T}} {\bf A}_k {\bf C}_k {\bf A}_k^{\mathrm{T}}$,
  and ${\bf L} \in \mathbb{R}^{n \times n}$ is the cotangent-weights Laplacian up to a
  constant scale factor.
Plugging in the linear blend skinning formula $\tilde{\bf V} = {\bf D}{\bf T}$ we obtain
\begin{align}
	2E({\bf V},\tilde{\bf V}) =& \trace({\bf T}^{\mathrm{T}} {\bf \tilde{L}}
    {\bf T}) - 2\trace({\bf R} {\bf \tilde{K}} {\bf T})
    + \trace({\bf V}^{\mathrm{T}} {\bf L} {\bf V}),
\end{align}
where ${\bf \tilde{L}} = {\bf {D}}^{\mathrm{T}}{\bf {L}}{\bf {D}}$ and ${\bf \tilde{K}} = {\bf {K}}{\bf {D}}$.
For more details about the above derivation, we refer the reader to \cite{jacobson2012fast}.


\section{Example-based blended transformations}
\noindent {\bf Overview.}
We now extend the framework described in the previous section for the case where multiple poses of the same shape are available.
We begin by expressing the deformed shape as a combination of atoms from a dictionary that is constructed from the linear blend skinning matrices of the given examples.
Then, we provide the details of 
 various energy terms to be minimized with respect to the unknown transformations ${\bf T}$ using the proposed model.
Next, we describe the nonlinear optimization process and its initialization, and
 conclude by discussing optional extensions that can be incorporated into the algorithm.

\subsection{ Dictionary Construction}
Suppose we are given $q$ reference meshes $\mathcal{M}_1, \mathcal{M}_2, \dots, \mathcal{M}_q$.
Let ${{\bf v}^{\ell}_1}, \dots, {\bf v}^{\ell}_n \in \mathbb{R}^{d}$ be the positions of vertices
 belonging to the reference mesh $ \mathcal{M}_\ell$, $\ell = 1, \dots,  q$,
 and let ${\bf V}_1, {\bf V}_2, \allowbreak \dots, {\bf V}_q \in \mathbb{R}^{n \times d}$
 be the matrices whose rows denote the positions of the corresponding vertices.
We are also given some $h$ linear constraints represented by the matrix
  ${\bf H} \in  \mathbb{R}^{h \times n}$,   such that
  ${\bf H}\tilde{\bf V} \approx {\bf Y}$, where ${\bf Y} \in \mathbb{R}^{h \times d}$
  is the value of these constraints for the target shape.
We  can define the linear constraints to be simply the coordinates of points on the mesh
 or use more refined measures such as the Laplacian coordinates
 \cite{alexa2003differential,lipman2004differential}, or a weighted average of
 some vertex positions, to constrain our non-rigid blended shape deformation.
Using this setup, we are interested in finding the positions of the target vertices
 as a result of a smooth transformation of the input meshes such that it
 approximately preserves local rigidity and satisfies the linear constraints
 up to a small error.

 \begin{figure*}[ht]
 	\begin{center}
 	\begin{overpic}[width=1.0\textwidth]{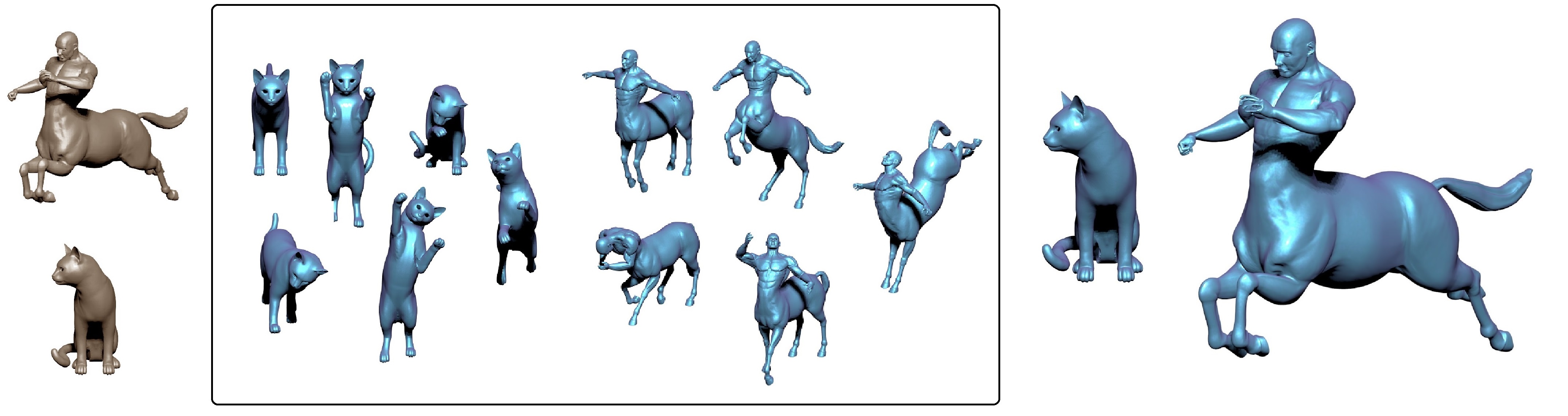}
 	\end{overpic}
 	\end{center}
 	\caption{\small
 Deformation using the example-based LBO dictionary.
 The left portion of the figure shows the cat and centaur \emph{ground-truth}
  target shapes (colored in gray).
 On the right we show the near perfect representation of these target shapes by a
  linear combination of the dictionary's atoms.
 In this case, the weighting functions are the eigenfunctions of the
  Laplace-Beltrami operator that correspond to the lowest $15$ eigenvalues.
 The exemplar shapes that are used to extract the example-based LBO dictionaries
  for representing the shapes are shown inside the box.
 }
 	\label{fig:examples}
 \end{figure*}


 \vspace{.5em}
 \noindent {\bf Example-based dictionary.}
 \vspace{.5em}
 Given $m$ real-valued weight functions $w_j$, $j=1 ,\dots, m$,
  we propose the example-based representation of the positions of the target vertices
  to be a combination of the linear blend skinning deformations of each given reference mesh
 \begin{align*}
  \tilde{\bf v}_i =& \sum_{\ell=1}^q \tilde{\bf v}^{\ell}_i,
 \end{align*}
 where
 \begin{align}
 	\tilde{\bf v}^{\ell}_i =& \sum_{j=1}^m w_{j,i} {\bf M}_j^\ell
      \begin{pmatrix}{\bf v}^{\ell}_i \\ 1\end{pmatrix}.
 \end{align}
 We can explicitly write the new vertex positions as
 \begin{align}
 \label{eq:v_def}
 \tilde{\bf v}_i =& \sum_{\ell=1}^q \sum_{j=1}^m w_{j,i} {\bf M}_j^\ell
  \begin{pmatrix}{\bf v}^{\ell}_i \\ 1\end{pmatrix} \cr
 	=& \sum_{\ell=1}^q\sum_{j=1}^m w_{j,i} {\bf \hat{M}}^\ell_j {\bf v}^{\ell}_i
      + \sum_{j=1}^m\sum_{\ell=1}^q w_{j,i} {\bf \bar{M}}^\ell_j ,
 \end{align}
  where ${\bf \hat{M}}^\ell_j \in \mathbb{R}^{d \times d}$ and
   ${\bf \bar{M}}^\ell_j \in \mathbb{R}^{d \times 1}$
    are sub-matrices of ${\bf M}_j^\ell$, such that ${\bf M}_j^\ell = \begin{pmatrix}
    {\bf \hat{M}}^\ell_j , & {\bf \bar{M}}^\ell_j\end{pmatrix}$.
 This formula can be equivalently expressed in the standard matrix form by
 \begin{align*}
 \tilde{\bf V} =& {\bf D}{\bf T},
 \end{align*}
 where ${\bf D} \in \mathbb{R}^{n \times (1+qd)m}$ is the proposed dictionary of size $b = (1+qd)m$, that multiplies the examples' vertex positions ${\bf v}^{\ell}_i$ with the vertex weights $w_j(v_i)$, and ${\bf T} \in \mathbb{R}^{(1+qd)m \times d}$ stacks the matrices  ${\bf \hat{M}}^m_j$  and ${\bf \bar{M}}^m_j$
  in the following way
 \begin{align*}
 {\bf D} =& \begin{pmatrix} {\bf \bar{D}}, & {\bf \hat{D}}_1 & \dots & {\bf \hat{D}}_q \end{pmatrix}, \cr
 {\bf T} =& \begin{pmatrix} {\bf \bar{T}^{\mathrm{T}}}, & {\bf \hat{T}}_{1}^{{\mathrm{T}}} & \hdots & {\bf \hat{T}}_{q}^{{\mathrm{T}}}\end{pmatrix}^{\mathrm{T}},
 \end{align*}
 where
 \begin{align*}
 {\bf \hat{D}}_\ell =& \begin{pmatrix} w_{1,1}  {{\bf v}_1^{\ell}}^{\mathrm{T}} & \dots  &w_{m,1}  {{\bf v}_1^{\ell}}^{\mathrm{T}} \\ \vdots & \ddots & \vdots \\ w_{1,n}  {{\bf v}_n^{\ell}}^{\mathrm{T}} & \dots  &w_{m,n} {{\bf v}_n^{\ell}}^{\mathrm{T}} \end{pmatrix}, \cr
 {\bf \hat{T}}_\ell =& \begin{pmatrix} {\bf \hat{M}}_1^{\ell} & \hdots & {\bf \hat{M}}_m^{\ell}\end{pmatrix}^{\mathrm{T}},
 \end{align*}
 and
 \begin{align*}
 {\bf \bar{D}} =& \begin{pmatrix} w_{1,1}   & \dots  &w_{m,1}  \\ \vdots & \ddots & \vdots \\
     w_{1,n}  & \dots  &w_{m,n} \end{pmatrix}, \\
    {\bf \bar{T}} =& \begin{pmatrix} \sum\limits_{\ell=1}^q{\bf \bar{M}}_1^{\ell} &
    \hdots & \sum\limits_{\ell=1}^q{\bf \bar{M}}_m^{\ell} \end{pmatrix}^{\mathrm{T}}.
 \end{align*}


 \vspace{.5em}
 \noindent {\bf Weighting functions.}
 There are many ways to choose weighting functions.
 One is to consider the weights of bones like in the
  standard linear blend skinning model.
 In that case, we name the constructed dictionary as the
  {\em example-based skeleton dictionary}.
 When there is no significant underlying skeletal structure,
  we suggest to use the first $m$ eigenfunctions of the Laplace-Beltrami operator (LBO) \cite{dey2012eigen,bouaziz2013online}. This choice of a dictionary is useful, for example, when handling  facial expressions, for analyzing internal organs in volumetric medical imaging applications, or for deforming non-rigid objects such as an octopus. The eigendecomposition of the LBO consists of non-negative eigenvalues
   $0 = \lambda_0 < \lambda_1 < \dots < \lambda_i < \dots$,
  with corresponding eigenfunctions
   \mbox{$\Phi \equiv \{\phi_0, \phi_1, \dots, \phi_i, \dots\}$},
   that can be considered as an orthonormal basis.
 We refer to this dictionary as the {\em example-based LBO dictionary}.


 \subsection{Nonlinear Energy Terms}
 \vspace{.5em}


 \noindent {\bf Linear constraints.}
 \vspace{.5em}
 The energy of the $h$ linear constraints can be calculated by
 \begin{align}
 \label{eq:lc}
  2E_{\text{lc}}(\tilde{\bf V})
              =& \norm{{\bf H} \tilde{\bf V} - {\bf Y}}_2^2
                = \norm{{\bf H} {\bf D} {\bf T} - {\bf Y}}_2^2
                = \norm{{\bf X}  {\bf T} - {\bf Y}}_2^2 \cr
              =& \trace{({\bf T}^{\mathrm{T}}{\bf X}^{\mathrm{T}} {\bf X}{\bf T})}
                  - 2\trace{({\bf Y}^{\mathrm{T}}{\bf X} {\bf T})}
                  + \trace{({\bf Y}^{\mathrm{T}}{\bf Y})},
 \end{align}
  where ${\bf X} = {\bf H} {\bf D}$.

 \vspace{.5em}
 \noindent {\bf Smoothness energy.}
 Let ${\bf  v}_{i,k},\: k \in \{1,\dots, d\}$ be the $k^{\mbox{th}}$
  coordinate of
  the vertex position ${\bf  v}_i$. Notice from Equation (\ref{eq:v_def}),
  that the amount of influence of ${\bf  v}^{\ell}_{i,k}$ on $\tilde{\bf  v}_{i,\tilde{k}}$
  is  some linear combination of $w_{j}(v_i)$, $j=1 ,\dots, m$.
 Following the same reasoning as in \cite{jacobson2011bounded},
  we search for a smooth variation of this influence, for example,
   one that minimizes the Laplacian energy $ \dfrac{1}{2} \int_{\cal M} \Delta(\cdot) ^2 da$ of this linear combination, where $da$ is an area element on the surface ${\cal M}$ of our shape.
 For the special case where the weights are the LBO eigenfunctions,
 the sum of all smoothness energy terms can be expressed as
 \begin{align}
 \label{eq:sm}
 E_{\text{sm}} =& \dfrac{1}{2} \trace({\bf T}^{\mathrm{T}}{\bf \Lambda} {\bf T}),
 \end{align}
 where ${\bf \Lambda}$ is a diagonal matrix. The values of the diagonal are the squares of the eigenvalues of the respective eignefunctions.
 Thus, in this case, the smoothness energy amounts to a
  simple quadratic regularization term.
 Note, that when the weighting functions are chosen in a different way,
  the smoothness energy expression is a bit more involved.


 \vspace{.5em}
 \noindent {\bf Scaling.}
 In some applications, there is a scale difference between the example shapes and the linear constraints.
 To compensate for such a discrepancy, we introduce a scaling factor $\alpha$ into the ARAP energy.
 It reflects the ratio between the reference shape and the deformed one, in the following manner,
 \begin{align}
 E_{\text{sc}}({\bf V},\tilde{\bf V})
    =& \cfrac{1}{2}\sum\limits_{k=1}^r\sum\limits_{(i,j)\in\mathcal{E}_k} c_{ijk}
     \norm{(\tilde{\bf v}_i-\tilde{\bf v}_j)-\alpha{\bf R}_k({\bf v}_i-{\bf v}_j)}^2.
 \end{align}
 Hence, the ARAP energy with the global scale factor reads
 \begin{align}
 \label{eq:arap}
 	2E_{\text{sc}}({\bf V},\tilde{\bf V}) = \trace({\bf T}^{\mathrm{T}} {\bf \tilde{L}} {\bf T}) - 2\alpha\trace({\bf R} {\bf \tilde{K}} {\bf T}) + \alpha^2\trace({\bf V}^{\mathrm{T}} {\bf L} {\bf V}).
 \end{align}


 \vspace{.5em}
 \noindent {\bf Average ARAP energy.}
 \vspace{.5em}
 One way to define an example-based energy functional is by taking the average between
  all as-rigid-as-possible energies, namely,
 \begin{align}
 \label{eq:eave}
 	E_{\text{av}} =& \dfrac{1}{q}\sum_{\ell=1}^q E_{\text{sc}}({\bf V}_\ell,\tilde{\bf V}),
 \end{align}
  with the additional linear constraints and the smoothness energies,
 \begin{align}
 	E_{\text{total}}(\tilde{\bf V}) =& E_{\text{av}}(\tilde{\bf V}) + \beta_{\text{lc}}E_{\text{lc}}(\tilde{\bf V}) + \beta_{\text{sm}}E_{\text{sm}}(\tilde{\bf V}),
 \end{align}
  where $\beta_{\text{lc}}$, $\beta_{\text{sm}}$, are some tuning parameters that control the
  importance of the linear constraints and the smoothness term.
 We can simplify this expression, plugging in \hbox{Equations (\ref{eq:sm})}, (\ref{eq:lc}) and (\ref{eq:arap})
 \begin{align}
 \label{eq:tot}
 	2E_{\text{total}}(\tilde{\bf V})  =&
     \trace({\bf T}^{\mathrm{T}} {\bf \tilde{L}} {\bf T}) \cr
      & - \cfrac{1}{q}  \sum_{\ell=1}^q (2\alpha \trace({\bf R}_\ell
      {\bf \tilde{K}}_\ell {\bf T}) + \alpha^2
           \trace({\bf V}_\ell^{\mathrm{T}} {\bf L} {\bf V}_\ell) )
      \cr
      &+ \beta_{\text{lc}}\trace{({\bf T}^{\mathrm{T}}{\bf X}^{\mathrm{T}} {\bf X}{\bf T})}
      -2 \beta_{\text{lc}}\trace{({\bf Y}^{\mathrm{T}}{\bf X} {\bf T})}  \cr
     &+\beta_{\text{lc}}\trace{({\bf Y}^{\mathrm{T}}{\bf Y})}
      + \beta_{\text{sm}} \trace({\bf T}^{\mathrm{T}}{\bf \Lambda} {\bf T}).
 \end{align}

 \begin{figure*}[t]
 	\begin{center}
 	\begin{overpic}[width=1.0\textwidth]{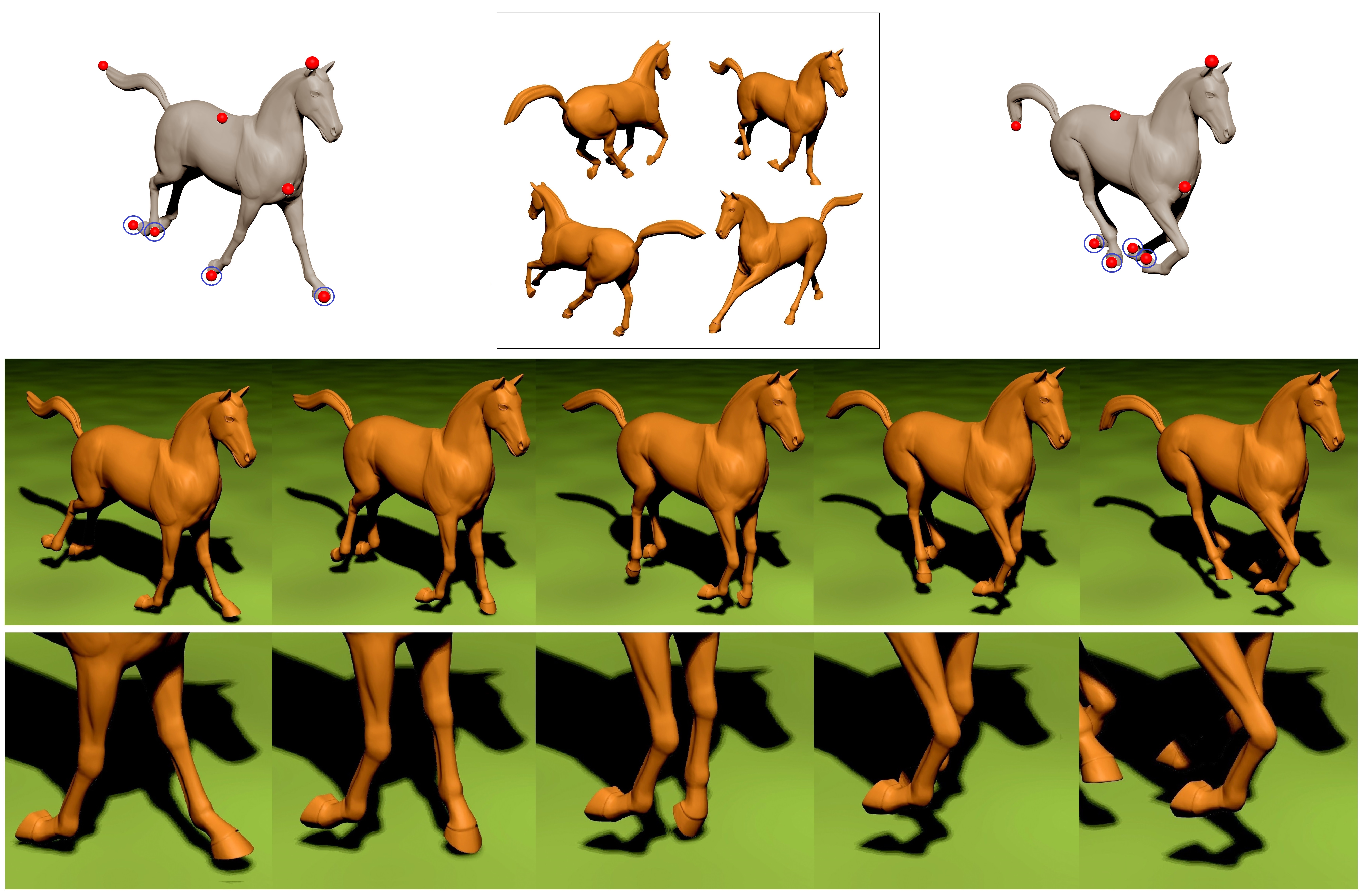}
 	\end{overpic}
 	\end{center}
 	\caption{\small Automatic feature point correspondence and shape interpolation. The four examples of a horse (top middle) and the two sets of
     vertex positions
      (top left, top right) were used to generate a sequence of frames.
      Correspondence of the points
       on the four legs (circled in blue) was detected by minimizing the
       example-based deformation energy
        for all permissible correspondences. The example-based deformations (bottom left and right)
         were then interpolated at four times the original frame rate to produce the movie sequence (bottom).
         }
 	\label{fig:p_corr}
 \end{figure*}


 \vspace{.5em}
 \noindent {\bf Minimal ARAP energy.}
 \vspace{.5em}
 Another way to define an example-based energy functional is to find the minimal
   ARAP energy between the deformed mesh and each of the input meshes separately,
 \begin{align}
 \label{eq:emin_all}
 E_{\text{mn}}(\tilde{\bf V})  =& \min_\ell E_{\ell}(\tilde{\bf V}, {\bf T}_\ell),
 \end{align}
 where
 \begin{align}
 \label{eq:emin}
 	E_{\ell}(\tilde{\bf V}, {\bf T}_\ell) =& E_{\text{sc}}({\bf V}_\ell,\tilde{\bf V}) + \beta_{\text{lc}}E_{\text{lc}}(\tilde{\bf V}) + \beta_{\text{sm}}E_{\text{sm}}(\tilde{\bf V}). \end{align}
 This can be expressed as
 \begin{align}
    2E_{\ell}(\tilde{\bf V}, {\bf T}_\ell) =&
         \trace({\bf T}_\ell^{\mathrm{T}} {\bf \tilde{L}} {\bf T}_\ell)
          - 2\alpha_\ell\trace({\bf R}_\ell {\bf \tilde{K}}_\ell {\bf T}_\ell) \cr
    &+ \alpha_\ell^2\trace({\bf V}_\ell^{\mathrm{T}} {\bf L} {\bf V}_\ell)
      + \beta_{\text{lc}}\trace{({\bf T}_\ell^{\mathrm{T}}{\bf X}^{\mathrm{T}} {\bf X}{\bf T}_\ell)}  \cr
      &- 2\beta_{\text{lc}}\trace{({\bf Y}^{\mathrm{T}}{\bf X} {\bf T}_\ell)}
         +\beta_{\text{lc}}\trace{({\bf Y}^{\mathrm{T}}{\bf Y})} \cr
         & + \beta_{\text{sm}} \trace({\bf T}_\ell^{\mathrm{T}}{\bf \Lambda} {\bf T}_\ell).
 \end{align}


 \subsection{Optimization}
 To minimize the energy $E_{\text{total}}(\tilde{\bf V})$ and find the local rotations
  ${\bf R}_\ell,\,\ell=1,\dots, q$, the global scale factor $\alpha$ and the transformations
  ${\bf T}$, we follow the local-global approach
   of \cite{sorkine2007rigid} with an additional step to
   find the global scale $\alpha$.
 First we fix ${\bf T}$and $\alpha$ and solve for ${\bf R}_\ell$
  (local step).
 Then, we find $\alpha$ by fixing ${\bf T}$, ${\bf R}_\ell$ (scale step).
 Finally, we fix ${\bf R}_\ell$ and $\alpha$, and solve for ${\bf T}$ (global step).
 \vspace{.5em}
 {\em Local step.} For fixed  $\alpha$ and ${\bf T}$,
  maximizing $\trace({\bf R}_\ell {\bf S}_\ell)$,  $\ell=1,\dots, q$,
  where ${\bf S}_\ell = {\bf \tilde{K}}_\ell {\bf T}_\ell$ is constant, amounts to maximizing $\trace{({\bf R}_{\ell,k} {\bf S}_{\ell,k})}$,  $k=1,\dots, r$,
  which is obtained by taking
  ${\bf R}_{\ell,k} = {\bf \Psi}^{\mathrm{T}}_{\ell,k} {\bf \Phi}^{\mathrm{T}}_{\ell,k}$,
 where $$ {\bf S}_{\ell,k} = {\bf \Phi}_{\ell,k} {\bf \Sigma}_{\ell,k}{\bf \Psi}_{\ell,k}$$
  is given by the singular value decomposition of $ {\bf S}_{\ell,k}$.

 \vspace{.5em}
 {\em Scale step.} For fixed ${\bf T}$ and ${\bf R}_\ell$, $\ell=1,\dots, q$,
  we can differentiate by $\alpha$
 \begin{align}
 	\dfrac{\partial E_{\text{total}}}{\partial \alpha} =&  -\dfrac{1}{q}
     \sum_{\ell=1}^q (\trace({\bf R}_\ell {\bf \tilde{K}}_\ell {\bf T}))
     + \alpha\trace({\bf V}_\ell^{\mathrm{T}} {\bf L} {\bf V}_\ell).
 \end{align}
 Setting the derivative to zero, we get
 \begin{align}
 	\alpha =&  \dfrac{1}{q} \sum_{\ell=1}^q
     (\trace({\bf R}_\ell {\bf \tilde{K}}_\ell {\bf T}))/
     \trace({\bf V}_\ell^{\mathrm{T}} {\bf L} {\bf V}_\ell).
 \end{align}

 \vspace{.5em}
 {\em Global step.} For fixed $\alpha$ and ${\bf R}_\ell$, $\ell=1,\dots, q$, we differentiate
  $ E_{\text{total}}$
 \begin{align}
 	\dfrac{\partial E_{\text{total}}}{\partial {\bf T}}
      =& \dfrac{1}{q} \sum_{\ell=1}^q ({\bf \tilde{L}} {\bf T} -
       \alpha{\bf \tilde{K}}_\ell^{\mathrm{T}}{\bf R}_\ell^{\mathrm{T}}) \cr
       & + \beta_{\text{lc}}({\bf X}^{\mathrm{T}} {\bf X}{\bf T} - {\bf X}^{\mathrm{T}}{\bf Y})
        + \beta_{\text{sm}} {\bf \Lambda} {\bf T} \cr
      =& ({\bf \tilde{L}}+\beta_{\text{lc}}{\bf X}^{\mathrm{T}} {\bf X}
          + \beta_{\text{sm}} {\bf \Lambda}) {\bf T}
           - \beta_{\text{lc}} {\bf X}^{\mathrm{T}}{\bf Y}
           \cr &
           - \dfrac{\alpha}{q}\sum_{\ell=1}^q
           {\bf \tilde{K}}_\ell^{\mathrm{T}}{\bf R}_\ell^{\mathrm{T}}.
 \end{align}
 Setting these derivatives to zero, we obtain
 \begin{align}
 ({\bf \tilde{L}}+\beta_{\text{lc}}{\bf X}^{\mathrm{T}} {\bf X} + \beta_{\text{sm}}
   {\bf \Lambda}) {\bf T}
   =  \beta_{\text{lc}}{\bf X}^{\mathrm{T}}{\bf Y} + \dfrac{\alpha}{q} \sum_{\ell=1}^q
       {\bf \tilde{K}}_\ell^{\mathrm{T}}{\bf R}_\ell^{\mathrm{T}}.
 \end{align}
 Let us define ${\bf \Gamma} = ({\bf \tilde{L}}+\beta_{\text{lc}}{\bf X}^{\mathrm{T}} {\bf X} + \beta_{\text{sm}} {\bf \Lambda})$. Then, we can solve for ${\bf T}$ by precomputing the Cholesky factorization of ${\bf \Gamma}$
 \begin{align}
 {\bf T} =&  {\bf \Gamma}^{-1}\big(\beta_{\text{lc}}{\bf X}^{\mathrm{T}}{\bf Y}
   + \dfrac{\alpha}{q}
    \sum_{\ell=1}^q {\bf \tilde{K}}_\ell^{\mathrm{T}}{\bf R}_\ell^{\mathrm{T}}\big).
 \end{align}

 As for optimizing the minimal ARAP energy $E_{\text{mn}}(\tilde{\bf V})$, in the local step we find each set of rotations ${\bf R}_\ell$ by maximizing $\trace({\bf R}_\ell {\bf S}_\ell)$, where
 ${\bf S_\ell} = \tilde{\bf K}_\ell{\bf T_\ell}$.
 We then find the global scale factor relative to each reference shape
 \begin{align*}
 	\alpha_\ell =&
     \trace({\bf R}_\ell {\bf \tilde{K}}_\ell
       {\bf T_\ell})/\trace({\bf V}_\ell^{\mathrm{T}} {\bf L} {\bf V}_\ell).
 \end{align*}
 In the global step we calculate the respective blended transformations ${\bf T_\ell}$, by
 \begin{align*}
 {\bf T}_\ell =&  {\bf \Gamma}^{-1}\big(\beta_{\text{lc}}{\bf X}^{\mathrm{T}}{\bf Y} + \alpha_\ell{\bf \tilde{K}}_\ell^{\mathrm{T}}{\bf R}_\ell^{\mathrm{T}}\big).
 \end{align*}
 Then, we calculate the minimal energy $E_\ell(\tilde{\bf V}, {\bf T}_\ell)$,
     $\ell = 1,\dots ,q$ of \hbox{Equation (\ref{eq:emin})}.


 \vspace{.5em}
 \noindent {\bf Initial transformations.}
 \vspace{.5em}
 In the first global step, there are no rotation matrices that can be used. Hence, the energy that we need to minimize is
 \begin{align}
 \label{eq:en_init}
 	2E_{\text{init}}(\tilde{\bf V}) =& \beta_{\text{lc}} \norm{{\bf X} {\bf T}
         - {\bf Y}}^2 + \beta_{\text{sm}} {\bf T}^{\mathrm{T}}{\bf \Lambda} {\bf T} \cr
       =&  \beta_{\text{lc}}\trace({\bf T}^{\mathrm{T}}{\bf X}^{\mathrm{T}} {\bf X}{\bf T})
             -2\beta_{\text{lc}}\trace({\bf Y}^{\mathrm{T}}{\bf X} {\bf T})  \cr
        & +\beta_{\text{lc}}\trace({\bf Y}^{\mathrm{T}}{\bf Y})
          + \beta_{\text{sm}} \trace({\bf T}^{\mathrm{T}}{\bf \Lambda} {\bf T}).
 \end{align}
 We readily have,
 \begin{align}
 	\dfrac{\partial E_{\text{init}}}{\partial {\bf T}}
     =&
      \beta_{\text{lc}}({\bf X}^{\mathrm{T}} {\bf X}{\bf T}
       - {\bf X}^{\mathrm{T}}{\bf Y})
       + \beta_{\text{sm}} {\bf \Lambda} {\bf T} \cr
   =& (\beta_{\text{lc}}{\bf X}^{\mathrm{T}} {\bf X}
    + \beta_{\text{sm}} {\bf \Lambda}) {\bf T}
    - \beta_{\text{lc}} {\bf X}^{\mathrm{T}}{\bf Y}.
 \end{align}
 Setting these derivatives to zero, we obtain
 \begin{align}
  {\bf T} =&   (\beta_{\text{lc}}{\bf X}^{\mathrm{T}} {\bf X} + \beta_{\text{sm}}
  {\bf \Lambda})^{-1}(\beta_{\text{lc}}{\bf X}^{\mathrm{T}}{\bf Y}).
 \end{align}


 \vspace{.5em}
 \noindent {\bf Sparse initial transformations.}
 A more robust initial transformation can be achieved by adding an $L_1$
    penalty to the energy given in \hbox{Equation (\ref{eq:en_init})}
 \begin{align}
 \label{eq:en_init_sp}
 	E_{\text{sp}}(\tilde{\bf V}) = E_{\text{init}}(\tilde{\bf V})
     + \beta_{\text{sp}}\norm{{\bf T}}_{L_1}.
 \end{align}
 The effect of this additional penalty is that it makes the initial
    transformations sparse, which results in a deformation with less artifacts.
 The parameter $\beta_{\text{sp}}$ controls the amount of sparsity
   in the initial solution of ${\bf T}$.
 \hbox{Equation (\ref{eq:en_init_sp})} can be solved efficiently using
    the elastic net regression method \cite{zou2005regularization}.


 \subsection{Extensions}
 \vspace{.5em}
 \noindent {\bf Updating constraints.}
 It may happen that some of the $h$ linear constraints are unavailable
   due to noise or occlusions.
 This can be easily solved by deleting the appropriate rows
  of ${\bf X}$ and ${\bf Y}$ and efficiently updating the Cholesky factorization.


 \vspace{.5em}
 \noindent {\bf Dictionary reduction.}
 When the input meshes are similar to each other,
  the proposed example-based dictionary becomes redundant.
 The dictionary can be reduced considerably by clustering similar dictionary atoms.
 For this purpose, we use the k-medoids clustering algorithm \cite{kaufman1987clustering}.
 The advantage of k-medoids over k-means clustering is that each cluster center of the
  k-medoids procedure is represented by one of the original dictionary atoms.
 This makes the appearance of the deformed shape more plausible compared to using k-means
  clustering for dimensionality reduction.


 \vspace{.5em}
 \noindent {\bf Change of dictionaries.}
 It is sometimes useful to work with two different dictionaries. In that case, the representations of the mesh in these two subspaces can be converted from one to the other in a simple way.
 Suppose we are given the dictionaries ${\bf D}_1$, ${\bf D}_2$
  and a good approximation of the transformation ${\bf T}_1$.
 Then, the transformation ${\bf T}_2$ can be set to
 \begin{align}
 \label{eq:dictionary_change}
 	{\bf T}_2 =& ({\bf D}_2^{\mathrm{T}}{\bf D}_2)^{-1}{\bf D}_1^{\mathrm{T}}{\bf T}_1.
 \end{align}
 This is particularly useful when one wants to initialize the transformations using a low
  dimensional dictionary by applying
  Equation (\ref{eq:en_init_sp}), and then change
  to a richer dictionary for obtaining more refined transformations.

 \begin{figure}[ht]
 	\begin{center}
 	\begin{overpic}[width=1\columnwidth]{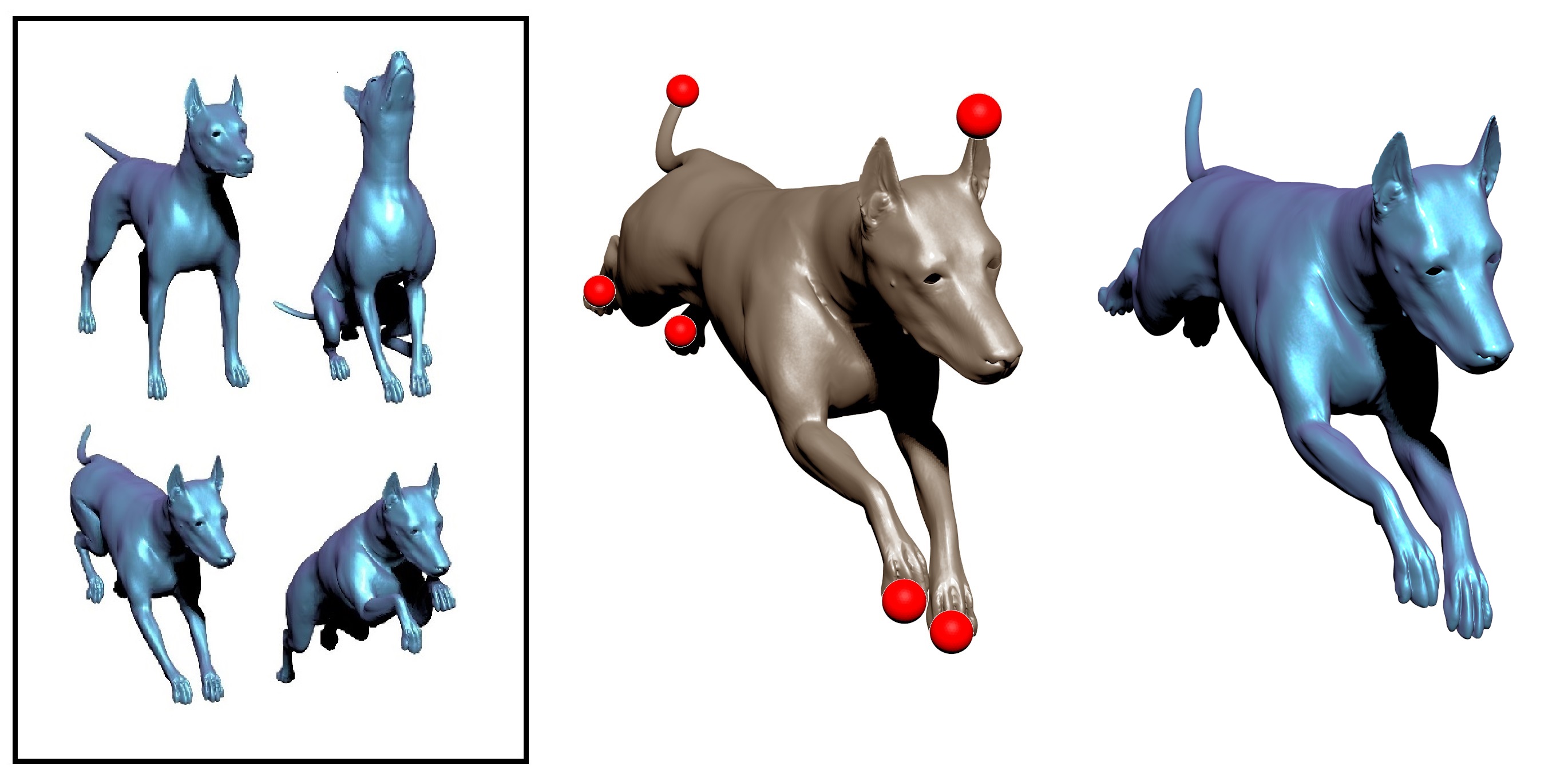}\end{overpic}
 	\end{center}
 	\caption{\small The four dog shapes are used as examples for our method (left).
         The deformed shape (right) is found from the vertex positions (middle).
         In this case the deformed shape is 50\% larger than the reference ones.}
 	\label{fig:eb_def_dog}
 \end{figure}

 \begin{figure}[ht]
 	\begin{center}
 	\begin{overpic}[width=\columnwidth]{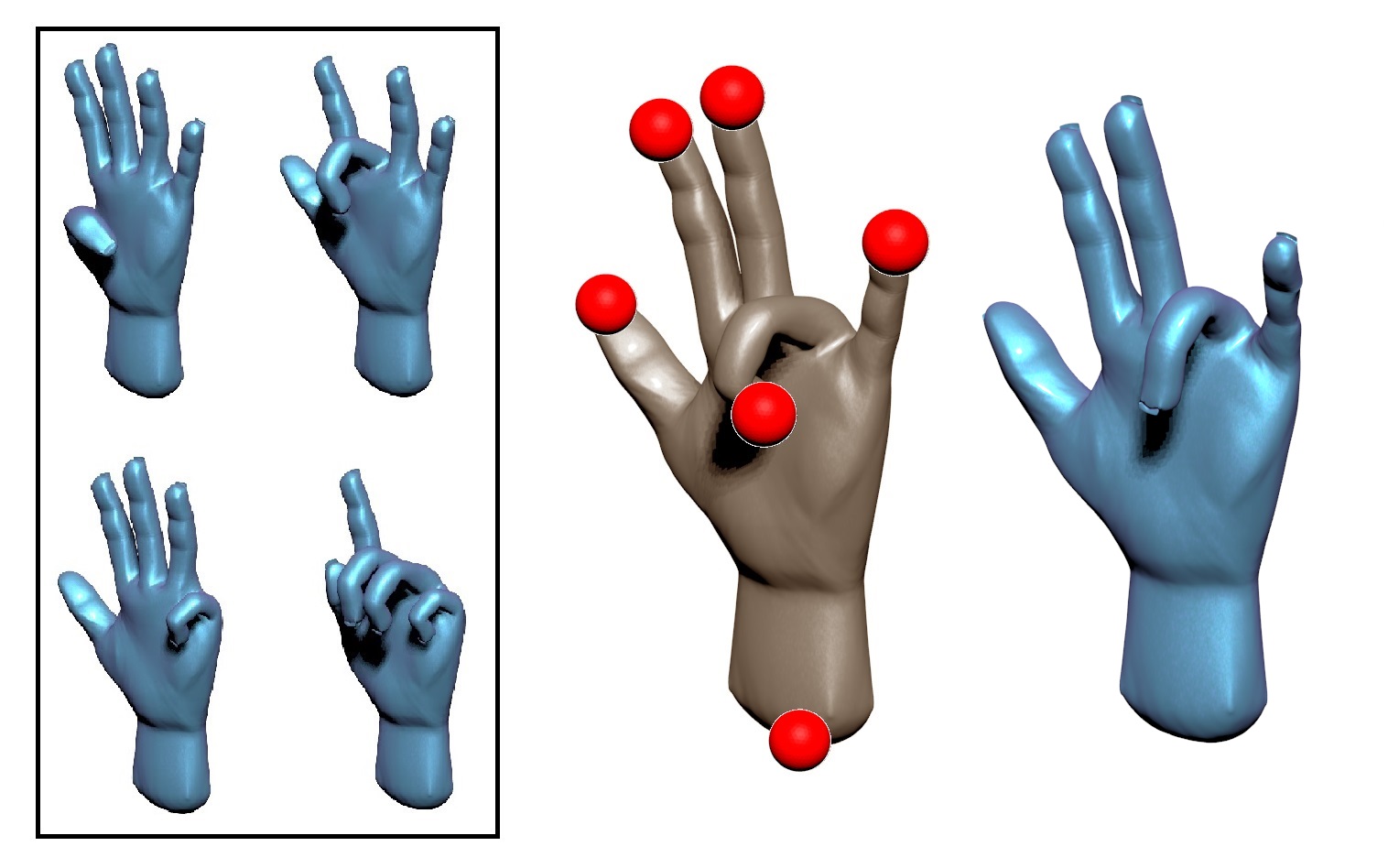}\end{overpic}
 	\end{center}
 	\caption{\small Example-based deformed shapes from few vertex positions of a hand shape.}
 	\label{fig:eb_def}
 \end{figure}

 \begin{table}[ht]
 \renewcommand{\arraystretch}{1.1}
 \centering
 \caption {Model parameters and performance (in milliseconds). }
 \begin{tabular}{|| c | c | c | c | c | c | c |c ||}
 \hline
  &  \multicolumn{2}{c|}{Input mesh} & \multicolumn{3}{c|}{Model} & \multicolumn{2}{c||}{Performance}\\
 \hline
 class & $n$ & $f$ & $r$ & $b$ & $q$ & $1$-iter & full \\
 \hline
 \hline
 woman & 45659 & 91208 & 30 & 178 & 5 & 1.3 & 16.7 \\
 \hline
 centaur & 15768 & 31532 & 31 & 166 & 5 & 1.2 & 16.3 \\
 \hline
 wolf & 4344 & 8684 & 14 & 82 & 2 & 0.7 & 10.3 \\
 \hline
 dog & 25290 & 50528 & 25 & 136 & 4 & 1.0 & 13.9 \\
 \hline
 man & 52565 & 105028 & 31 & 235 & 9 & 1.9 & 23.2 \\
 \hline
 cat & 27894 & 55712 & 22 & 148 & 6 & 1.2 & 15.3 \\
 \hline
 hand & 2224 & 4424 & 18 & 163 & 7 & 1.2 & 15.5 \\
 \hline
 horse & 16843 & 8431 & 26 & 139 & 4 & 1.1 & 14.7 \\
 \hline
 \end{tabular}
 \label{tbl:params}
 \end{table}

 \section{Experimental results}

 \begin{table}[ht]
 \renewcommand{\arraystretch}{1.1}
 \centering
 \caption {Example-based dictionary. Maximal relative distortion (in percent) of the
   deformed shapes. }
 \begin{tabular}{|| c | c | c | c | c ||}
 \hline
 \hline
   & dict. & example-based & example-based & one \\
     $q$ & size & LBO dict. & skeleton dict. & example \\
 \hline
 \hline
 1 & 94 & 2.18 & 3.59 & 2.39 \\
 \hline
 2 & 139 & 1.97 & 2.50 & 2.10 \\
 \hline
 3 & 184 & 1.82 & 2.16 & 1.95 \\
 \hline
 4 & 218 & 1.72 & 1.89 & 1.88 \\
 \hline
 5 & 258 & 1.63 & 1.78 & 1.83 \\
 \hline
 6 & 293 & 1.56 & 1.68 & 1.77 \\
 \hline
 7 & 339 & 1.52 & 1.60 & 1.71 \\
 \hline
 8 & 382 & 1.51 & 1.50 & 1.65 \\
 \hline
 \hline
 \end{tabular}
 \label{tbl:max_err}
 \end{table}

 \begin{figure*}[t]
 	\begin{center}
 	\begin{overpic}[width=1.0\textwidth]{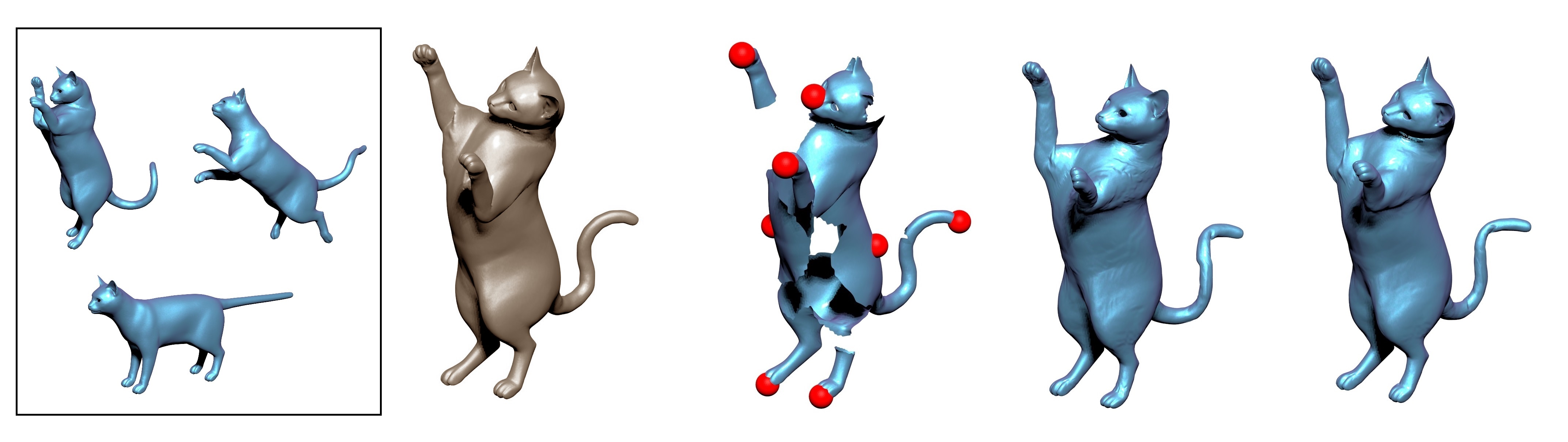}
 	\end{overpic}
 	\end{center}
 	\caption{\small Nonrigid ICP. Left to right: The three exemplar shapes, the uncorrupted and complete target shape, the acquired partial shape with eight known feature points (marked as red dots), the initial deformation using the feature points and the final deformation after applying the nonrigid ICP. }
 	\label{fig:nonrigid_icp}
 \end{figure*}

 \vspace{.5em}
 \noindent {\bf Implementation Considerations.}
 In our implementation we use $m \le 15$ eigenfunctions as the weighting functions for
  the example-based LBO dictionary.
 To support natural articulated shapes deformation, we construct the example-based skeleton dictionary.
 Its weights are generated using an automatic example-based skinning software package \cite{le2014robust}.
 These skeleton weights are also used to define the rotation clusters.
 After constructing the dictionary from our mesh examples, we decrease the size of the dictionary
  using the k-medoids clustering algorithm.
 This step typically reduces the size of the dictionary in half.

 The transformations are found in several steps.
 We begin by estimating the sparse initial transformations
  using Equation (\ref{eq:en_init_sp}).
 Typically, we start with $m \le 4$ eigenfunctions as the weighting functions.
 Then, we apply a two stage optimization procedure.
 In the first step we minimize the average ARAP energy of Equation (\ref{eq:tot}).
 This energy, although robust,  tends to smooth out some of the details of the shape.
 Therefore, in the second step we optimize the minimal ARAP energy of Equation
  (\ref{eq:emin_all}), that effectively selects  one example pose which seems to
  be closest to the target pose.
 After a few iterations, we apply Equation (\ref{eq:dictionary_change}),
  and change to a richer  dictionary that can reflect finer details of the shape.
 We construct this richer dictionary according to the properties of the subject we
  want to deform.
 For articulated shapes, we use the example-based skeleton dictionary.
 For non-articulated objects, we increase the number of eigenfunctions
  used to construct the example-based LBO dictionary.

 The algorithm was implemented in MATLAB with some optimizations in C++.
 We use the SVD routines provided by McAdams et al. \cite{mcadams2011computing}.
 All the experiments were executed on a 3.00 GHz Intel Core i7 machine with 32GB RAM.
 In Table \ref{tbl:params} we give the settings for different mesh classes  \cite{bronstein2008numerical,sumner2004deformation} and typical performance of the algorithm.
 For these settings the algorithm takes between 10 and 25 milliseconds.


 \vspace{.5em}
 \noindent {\bf Example-based dictionary.}
 The example-based dictionary spans natural deformations of a given shape with a small error.
 In \hbox{Figure \ref{fig:examples}} we show some examples of deformations created using
 the example-based LBO dictionary with $15$ eigenfunctions. The mesh parameters and number of example shapes used is as in Table \ref{tbl:params}.
 Observe, that there are no noticeable artifacts in these deformations.

 We note that the experiments indicate that
  the accuracy of the proposed model increases
  with the number of example shapes.
 For each shape in the database \cite{anguelov2005correlated},
  we found the closest deformed shape in the $L_2$ sense.
 We calculated the maximal Euclidean distortion between the deformed and the
  original shape and normalized it by the square root of the original shape's area.
 Then, we average this maximal distortion for all shapes.
 We notice, that the maximal distortion decreases as the number of example shapes grows.
 We also compare the example-based LBO and the skeleton dictionaries.
 Although quantitatively, the example-based LBO dictionary seems to perform better,
  our experience suggests that for shapes that have a well-defined skeleton,
  the example-based skeleton dictionary is more pleasing to the eye, as it captures
  the stiffness of the bones.
 Another conclusion is that using many examples improves the deformation accuracy.
 This can be seen by calculating the distortion of the deformed shape when the
  example-based LBO dictionary is constructed using one shape only,
  while keeping the number of dictionary atoms the same and without applying the dictionary
  reduction step.
 Table \ref{tbl:max_err} summarizes the results.


 \vspace{.5em}
 \noindent {\bf Example-based deformation from few vertex positions.}
 Perhaps, the most powerful application of our example-based framework
  is finding a naturally deformed shape from just a couple of vertex positions.
 In this scenario, we are given the positions of just a few points
  of a single depth image of a target shape.
 Given prior example shapes in different postures, we are able to faithfully and reliably
  reconstruct the target shape.
 In Figures  \ref{fig:eb_def_dog} and \ref{fig:eb_def}, we show reconstructed dog and hand,
  shapes from a small number of feature points.
 In these examples, the feature points were sampled in a scale different than
  that of the example-shapes by a factor of 1.5 (dog), and 0.7 (hand).


 \vspace{.5em}
 \noindent {\bf Automatic feature point correspondence.}
 The example-based deformation energy can be used to find correspondence between
  the example shapes and the given feature points \cite{zhang2008deformation}.
 Because our method does not rely on good initialization nor on many input points,
  it is ideal for such a purpose.
 For example, in \hbox{Figure \ref{fig:p_corr}}, we are given four reference shapes and eight feature points.
 In this demonstration, the correspondences of the four feature points that belong to
  each leg (circled in blue) are difficult to find.
 We can resolve this ambiguity by running our optimization algorithm for all $24$
  options of permissible correspondences, and calculate the example based energy
  of Equation (\ref{eq:emin_all}) for each.
 Then, the correspondence can be found by choosing the option that gave the minimal
  deformation energy.


\begin{figure*}[ht]
\centering
	\begin{overpic}[width=.9\columnwidth]{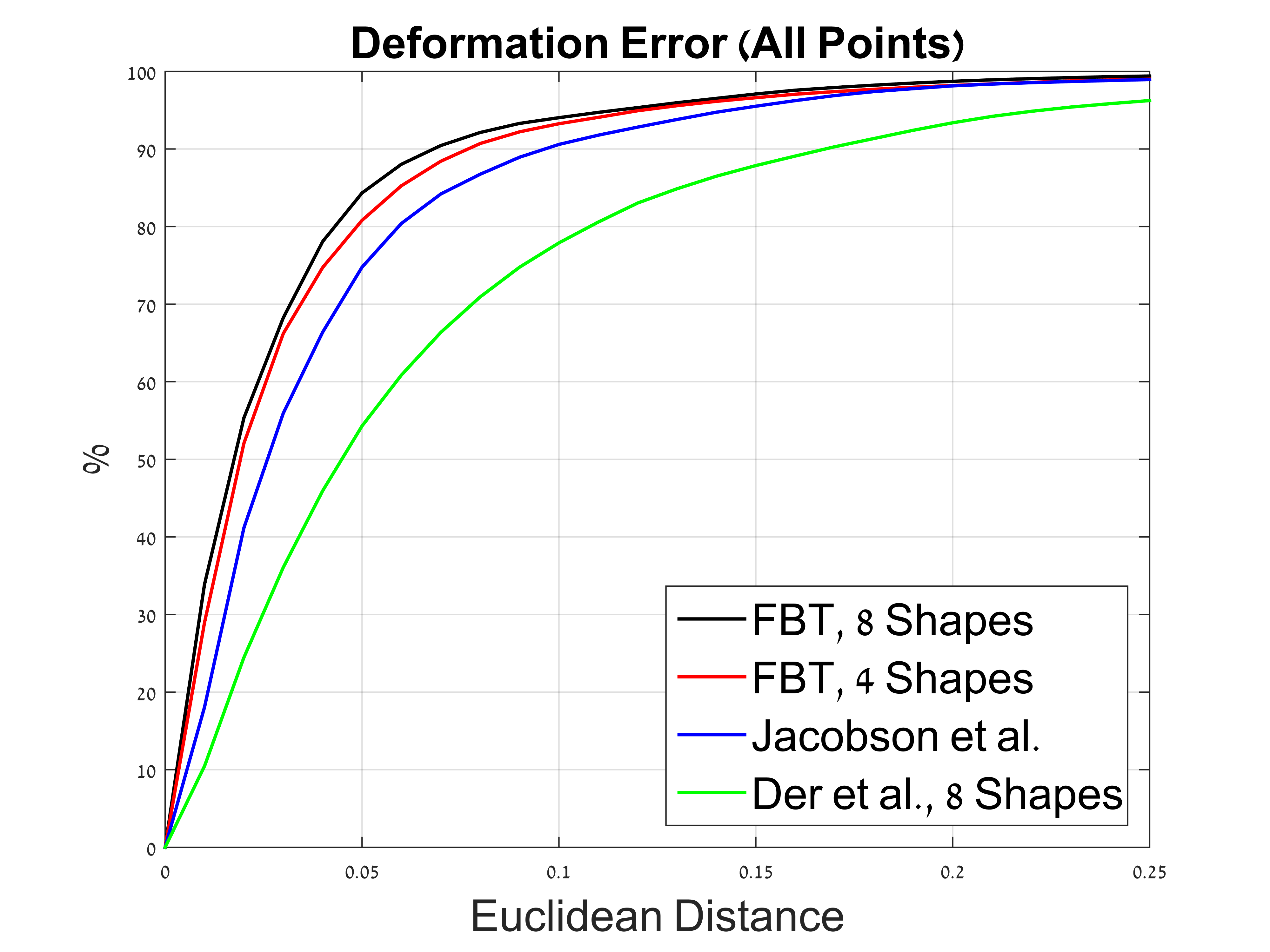}
	\end{overpic}
	\begin{overpic}[width=.9\columnwidth]{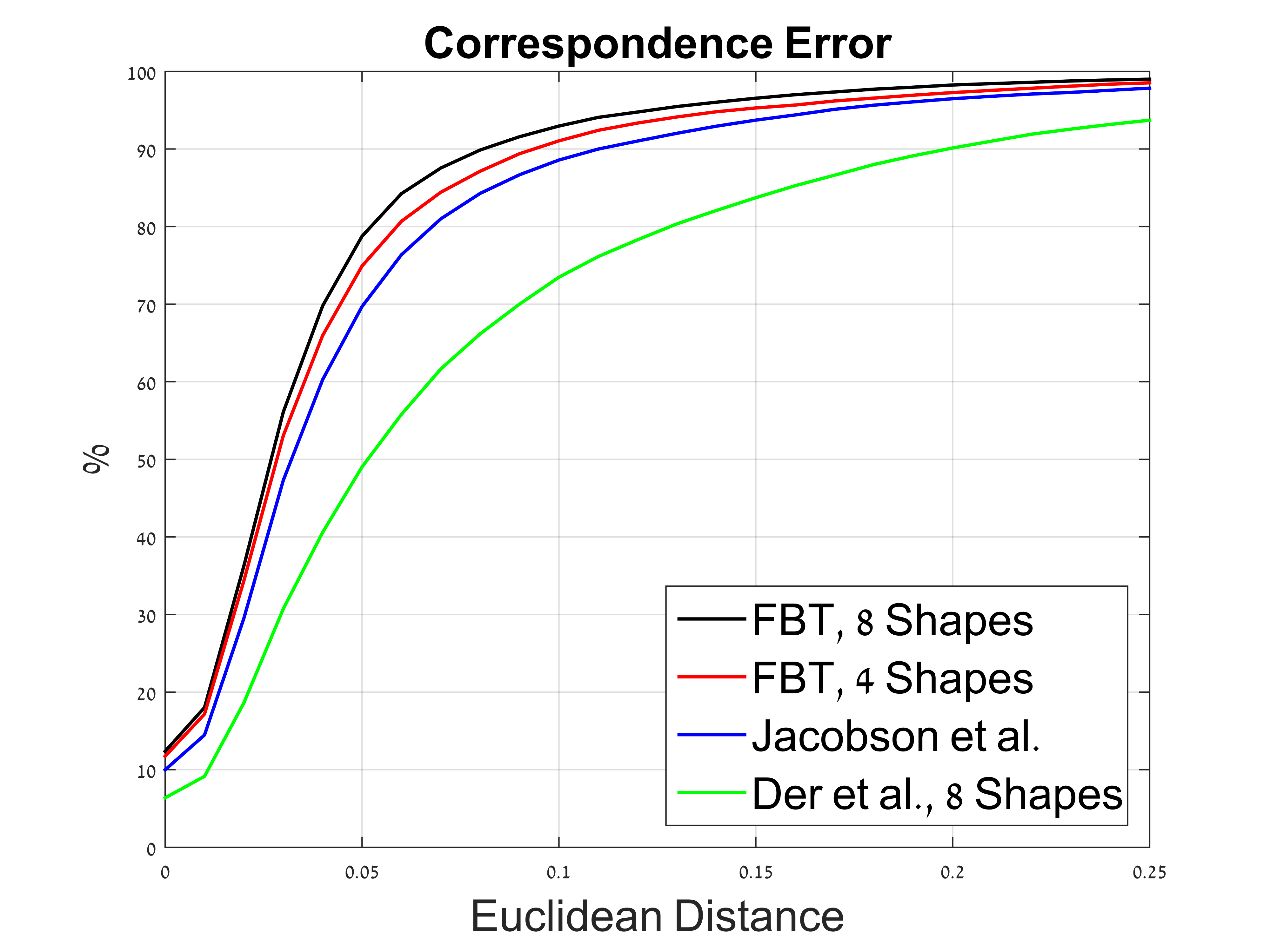}
	\end{overpic}
	\caption{\small Evaluation of the shape completion and registration procedure applied to shapes from the TOSCA database. }
	\label{fig:def_error_results}
\end{figure*}


\vspace{.5em}
\noindent {\bf Shape interpolation.}
A nice application that can easily be performed, is to interpolate between two deformed shapes.
In our setting, we are given two instances of positional constraints.
From these constraints we find two deformed shapes and their rotations.
Then, we are able to interpolate between these rotations.
To produce the new transformations, we apply one additional global-step.
\hbox{Figure \ref{fig:p_corr}} demonstrates an interpolation between two deformed shapes
 of a galloping horse.
Four example meshes are used as an input.
In the supplementary material we add a video of a galloping horse reconstructed from
 few feature points.
The video frames are interpolated by a factor of eight.
Based on the proposed ideas, we developed a computer program that automatically finds a natural deformed shape from
 a user's specified vertex locations and interpolates between the start pose and
 the final deformation of the shape, creating a smooth and intuitive
 motion of the shape. We provide a video that shows how this software is used to make an animation sequence of a moving person.


\noindent {\bf Nonrigid ICP.}
The blended transformations can be plugged into a simple nonrigid ICP framework \cite{allen2003space,amberg2007optimal,li2008global}. Nonrigid ICP registration alternates between finding pointwise correspondences and deforming one shape to best fit the other. Hence, we propose the following strategy. To find correspondences compare the vertex positions of all points and their surface normal vectors. In each iteration, we set new linear constraints according to the vertex positions of the obtained point-to-point correspondences, and apply our blended transformations method to wrap the nonrigid shapes while keeping the deformed shape inside the example manifold. We note that because the representation space is defined by the blended transformations it is suffice to match only a subset of points on the two shapes.

\FloatBarrier
\vspace{.5em}
\noindent {\bf Shape completion and registration.}
In many depth data acquisition scenarios, the acquired data consists of an incomplete, occluded and disconnected parts of a shape. Given some known feature points in those parts of the shape, we want to find the deformation that best fits the partial data and detect the pointwise mapping between acquired partial shape and the reference shapes. To this end, we propose a two step procedure. In the first step, the feature points are used to find an initial deformation. In the second step, the deformation is refined by applying a nonrigid ICP procedure.
Since our deformation technique is able to find a good approximation from just a few vertex positions it is ideal to be plugged into this procedure. Figure \ref{fig:nonrigid_icp} shows an example of partial data of a cat shape (middle) with some known feature points (marked in red). The initial deformation was found by applying the proposed Fast Blended Transformation algorithm using the known feature points (second from left). The final deformation was attained by applying the nonrigid ICP algorithm in conjunction with our blended transformations approach (left). Notice that the nonrigid ICP algorithm corrected the tilt of the cat's head.

We tested the proposed shape completion and registration procedure on shapes represented by triangulated meshes from the TOSCA database \cite{bronstein2008numerical}. We performed 50 random experiments with different example and target shapes. For each experiment, we were given eight reference shapes and one target shape for which some of its vertices were removed. We assume that the remaining shape includes some predefined parts that amount to more than 50\% of the shape's area. We farther assume that in these parts there are a number of identifiable feature points and that around each feature point, within a certain geodesic circle, no vertices were removed. In the test we performed the number of feature points was set to eight and the radius of the geodesic circle about each point was 15\% of the square-root of the shape's area.

We studied the performance of our approach with different number of example shapes. In our implementation, we set the initial linear constraints  to be the weighted average of the vertex positions in 10 different geodesic circles around each feature point. The weights of each vertex were proportional to its voronoi area. Using these linear constraints, we found an initial guess of the deformation. Then, we employed the nonrigid ICP algorithm for the rest of the mesh.
We also compared our results with the ones obtained by plugging in the deformation method proposed by Der et al. \cite{der2006inverse} into our shape completion procedure, using the same skeleton structure. This method applies an example-based deformation gradient model on the problem, and is computationally comparable to the proposed fast blended transformations algorithm. To achieve better results, we used a modified version of the deformation gradient model that supports soft constraints. For leveling the playing field, the automatic skeleton structure was found in the same way for all methods \cite{le2014robust}.

Figure \ref{fig:def_error_results} (left) compares the accuracy of the achieved deformations.
The distortion curves describe the percentage of surface points falling within a relative distance from the target mesh. For each shape, the Euclidean distance is normalized by the square root of the shape's area.
As for the partial registration, the distortion curves shown in Figure \ref{fig:def_error_results} (right) describe the percentage of correspondences that fall within
a relative Euclidean distance from what is assumed to be their true locations, similar to the protocol of \cite{kim2011blended}.
We see, that both the deformation quality and the correspondence accuracy increase with the number of reference shapes. This is expected, since as more example poses are introduced, the example-based dictionary better spans the space of natural deformations and we have more poses to compare against. We also notice that for these experiments, our deformation approach (even with one reference shape as in \cite{jacobson2012fast}) significantly outperforms the inverse kinematics method of Der et al. \cite{der2006inverse}. This can be explained by the fact that the reduced deformable model of Der et al. is based on explicit interpolation between the reference poses using deformation gradients. Apparently, this model needs a large number of reference poses to cover all the allowed isometric transformations. In contrast, our model implicitly finds the example manifold by a linear combination of the dictionary atoms and the ARAP energy. Hence, it needs far fewer examples.

\vspace{.5em}
\begin{figure}[ht]
	\begin{center}
	\fbox{
	\begin{overpic}[width=.95\columnwidth]{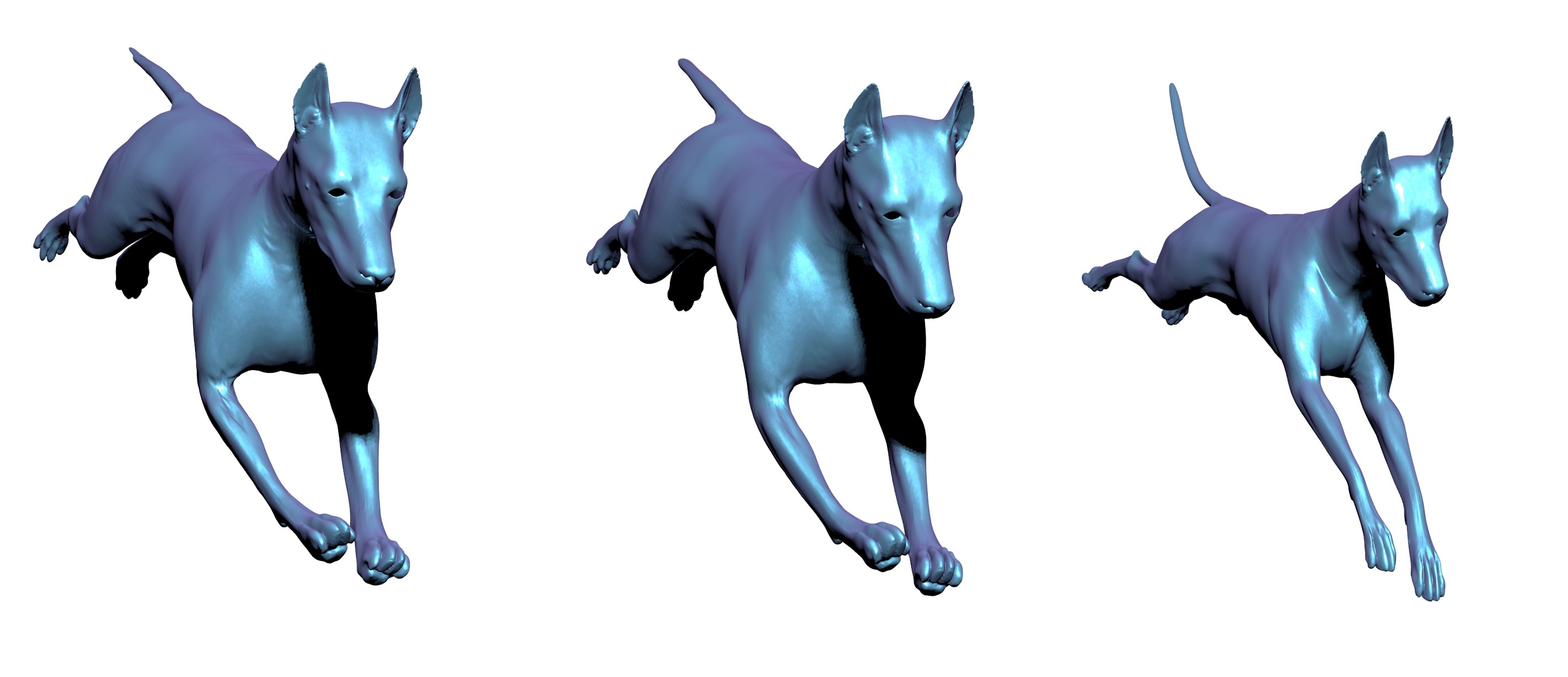}\put(5,1){\ref{en:base1}}\put(40,1){\ref{en:ref1}}\put(78,1){\ref{en:no_scale}}
	\end{overpic}
	}
	\end{center}
	\vspace{.1em}
	\begin{center}
	\fbox{
	\begin{overpic}[width=.95\columnwidth]{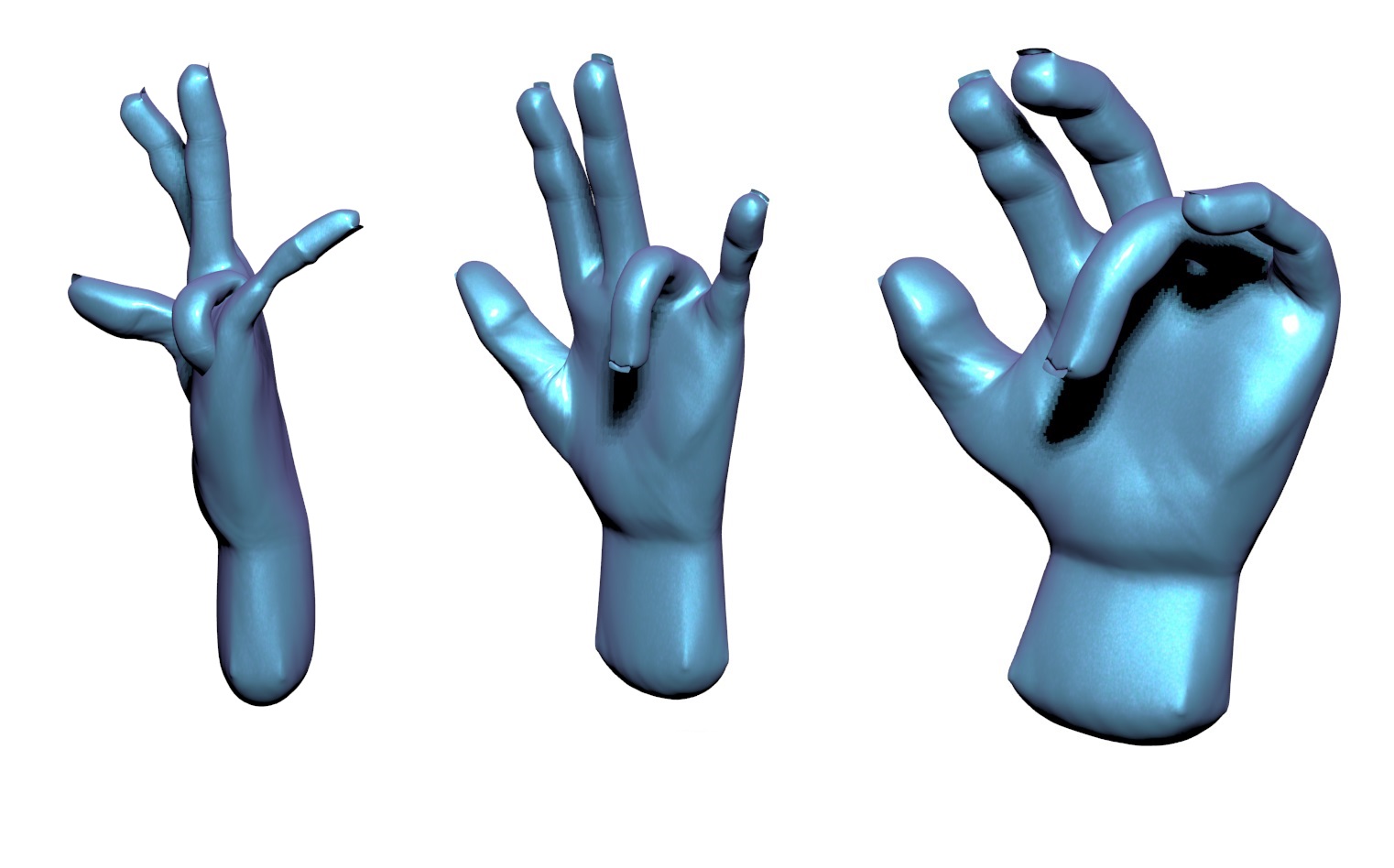}\put(5,1){\ref{en:base1}}\put(40,1){\ref{en:ref1}}\put(78,1){\ref{en:no_scale}}
	\end{overpic}
	}
	\end{center}
	\caption{\small Deformed shapes constructed by omitting some of the steps in the proposed
                    example-based framework. }
\vspace{.5em}
{\small
\begin{enumerate}[nosep,label=\emph{(\Alph*})]
\item \label{en:base1} Restricting the example-based dictionary to use one example.
\item \label{en:ref1} Comparing the deformed shape to only one reference shape.
\item \label{en:no_scale} Skipping the scale-step, by setting $\alpha = 1$.
\end{enumerate}
}
	\label{fig:eb_ver}
\end{figure}
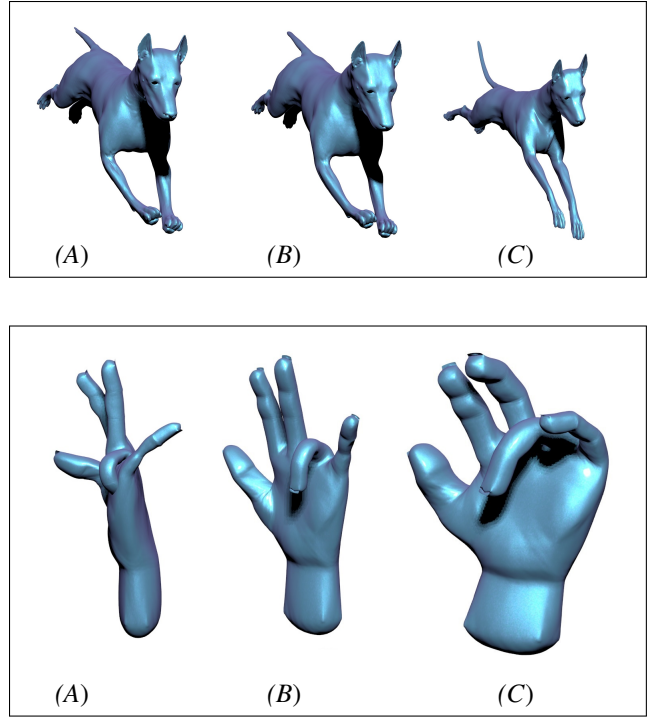
\FloatBarrier

\begin{table}[!ht]
\caption{List of Mathematical Symbols}
\renewcommand{\arraystretch}{1.5}
{
\begin{tabular}{|c|l|}
\cline{1-2}
\multicolumn{1}{|c|}{\textbf{Symbol}} & \multicolumn{1}{c|}{\textbf{Description}}\\
\hline
\hline
$\mathcal{M}$, $\tilde{\mathcal{M}}$ & reference and deformed meshes \\
\hline
$d$ & dimension of the shape \\
\hline
$n$ & number of vertices \\
\hline
$f$ & number of faces \\
\hline
$m$ & number of blending weight functions\\
\hline
$q$ & number of example shapes \\
\hline
$r$ & number of rotation clusters \\
\hline
${\bf v}_i^\ell$ & $i$th vertex of the $\ell$th shape \\
\hline
$w_{j,i}$ & $j$th blending weight function at the $i$th vertex\\
\hline
$\phi_{j}$ & $j$th eigenfunction of the Laplace-Beltrami operator \\
\hline
$\lambda_{j}$ &$j$th eigenvalue of the Laplace-Beltrami operator \\
\hline
$ {\bf M}_j^\ell$ & $j$th transformation of the $\ell$th shape\\
\hline
${\bf V}$,${\bf \widetilde{V}}$ & set of reference and deformed vertices  \\
\hline
${\bf D}$  & example-based dictionary\\
\hline
${\bf T}$ & transformation matrix\\
\hline
${\bf R}_k^{\ell}$ & $k$th rotation of the $\ell$th shape\\
\hline
$\mathcal{E}_k$ & set of vertices of the $k$th rotation cluster\\
\hline
$c_{ijk}$ & cotangent weight of the edge $(i,j)$ in the \\ & $k$th rotation cluster\\
\hline
$ {\bf H } $ & constraint sampling matrix \\
\hline
$ {\bf Y } $ & constraint matrix\\
\hline
$ {\bf \Lambda} $ & diagonal eigenvalue matrix\\
\hline
$E_{sm} $ & smoothness term\\
\hline
$E_{lc} $ & linear constraint term\\
\hline
$E_{sc} $ & scaled as-rigid-as-possible term\\
\hline
$E_{av} $ & averaged as-rigid-as-possible term\\
\hline
$\beta_{lc} $ & linear constraint weight\\
\hline
$\beta_{sm} $ & smoothness weight\\
\hline
$\alpha_{\ell} $ & scaling parameter of the $\ell$th shape\\
\hline
\end{tabular}
}
\label{tbl:variables}
\end{table}


\section{Discussion}
We tested some deficient versions of our example-based deformation framework.
In \hbox{Figure \ref{fig:eb_ver}} we show several examples of how these partial versions
 of the algorithm behave.
For comparison to the complete method see \hbox{Figures \ref{fig:eb_def_dog} and \ref{fig:eb_def}}.
We notice that the most important part of the proposed framework
 is the construction of the dictionary from multiple examples.
If only one example is used \ref{en:base1}, as in \cite{jacobson2012fast},
 the deformation algorithm fails when the shape has many degrees of freedom.

Although the method is robust and usually performs very well, some limitations and
 failures in particular cases do exist.
Despite the usually pleasing to the eye deformations of the proposed example-based approach,
 sometimes undesirable artifacts might occur.
This is the result of the collinearity between different dictionary atoms.
As for the performance of the algorithm, the deformation can be produced in real time
 but the algorithm cannot accommodate for video applications with many objects that
  need to be simultaneously deformed.
This problem can be solved by using the proposed algorithm only for objects for which
 a previous pose cannot be used for the initialization of the current one.
Another drawback is that if the example shapes do not incorporate enough information
 for extracting the right rotation clusters, then, the algorithm will ultimately fail.
Also, the current evaluation system does not prevent self-intersections.


\section{Conclusions}
We applied the concept of overcomplete dictionary  representation to the problem of shape deformation.
The proposed example-based deformation approach extends the subspace
 of physically-plausible deformations, while controlling the smoothness of the reconstructed mesh.
The blended transformations enable us to find a new pose from a small number
 of known feature points without any additional information.
It is well-suited for real-time applications as well as offline animation
 and analysis systems.
In the future, we plan to apply the proposed framework to various problems
 from the field of shape understanding, such as gesture recognition,
 registration of MRI images, and prior based object reconstruction
 from depth images.

 {\small
 \bibliographystyle{ieee}
 \bibliography{fast_blended_transformations}

\begin{thebibliography}{10}\itemsep=-1pt

\bibitem{alexa2003differential}
M.~Alexa.
\newblock Differential coordinates for local mesh morphing and deformation.
\newblock {\em The Visual Computer}, 19(2):105--114, 2003.

\bibitem{alexa2000rigid}
M.~Alexa, D.~Cohen-Or, and D.~Levin.
\newblock As-rigid-as-possible shape interpolation.
\newblock In {\em Proceedings of the 27th Annual Conference on Computer
  Graphics and Interactive Techniques}, SIGGRAPH '00, pages 157--164, New York,
  NY, USA, 2000. ACM Press/Addison-Wesley Publishing Co.

\bibitem{allen2003space}
B.~Allen, B.~Curless, and Z.~Popovi\'{c}.
\newblock The space of human body shapes: Reconstruction and parameterization
  from range scans.
\newblock {\em ACM Trans. Graph.}, 22(3):587--594, July 2003.

\bibitem{amberg2007optimal}
B.~Amberg, S.~Romdhani, and T.~Vetter.
\newblock Optimal step nonrigid icp algorithms for surface registration.
\newblock In {\em Computer Vision and Pattern Recognition, 2007. CVPR '07. IEEE
  Conference on}, pages 1--8, June 2007.

\bibitem{anguelov2005correlated}
D.~Anguelov, P.~Srinivasan, H.~cheung Pang, D.~Koller, S.~Thrun, and J.~Davis.
\newblock The correlated correspondence algorithm for unsupervised registration
  of nonrigid surfaces.
\newblock In L.~K. Saul, Y.~Weiss, and L.~Bottou, editors, {\em Advances in
  Neural Information Processing Systems 17}, pages 33--40. MIT Press, 2005.

\bibitem{bogo2014faust}
F.~Bogo, J.~Romero, M.~Loper, and M.~J. Black.
\newblock Faust: Dataset and evaluation for 3d mesh registration.
\newblock In {\em The IEEE Conference on Computer Vision and Pattern
  Recognition (CVPR)}, June 2014.

\bibitem{bouaziz2014projective}
S.~Bouaziz, S.~Martin, T.~Liu, L.~Kavan, and M.~Pauly.
\newblock Projective dynamics: Fusing constraint projections for fast
  simulation.
\newblock {\em ACM Trans. Graph.}, 33(4):154:1--154:11, July 2014.

\bibitem{bouaziz2013online}
S.~Bouaziz, Y.~Wang, and M.~Pauly.
\newblock Online modeling for realtime facial animation.
\newblock {\em ACM Transactions on Graphics (TOG)}, 32(4):40, 2013.

\bibitem{bronstein2008numerical}
A.~M. {B}ronstein, M.~M. {B}ronstein, and R.~{K}immel.
\newblock {\em {N}umerical geometry of non-rigid shapes}.
\newblock {S}pringer, 2008.

\bibitem{chao2010simple}
I.~Chao, U.~Pinkall, P.~Sanan, and P.~Schr\"{o}der.
\newblock A simple geometric model for elastic deformations.
\newblock In {\em ACM SIGGRAPH 2010 Papers}, SIGGRAPH '10, pages 38:1--38:6,
  New York, NY, USA, 2010. ACM.

\bibitem{der2006inverse}
K.~G. Der, R.~W. Sumner, and J.~Popovi\'{c}.
\newblock Inverse kinematics for reduced deformable models.
\newblock {\em ACM Trans. Graph.}, 25(3):1174--1179, July 2006.

\bibitem{dey2012eigen}
T.~K. Dey, P.~Ranjan, and Y.~Wang.
\newblock Eigen deformation of 3d models.
\newblock {\em The Visual Computer}, 28(6-8):585--595, 2012.

\bibitem{elad2010sparse}
M.~Elad.
\newblock {\em Sparse and redundant representations: {From} theory to
  applications in signal and image processing}.
\newblock Springer Science \& Business Media, 2010.

\bibitem{feng2008real}
W.-W. Feng, B.-U. Kim, and Y.~Yu.
\newblock Real-time data driven deformation using kernel canonical correlation
  analysis.
\newblock In {\em ACM SIGGRAPH 2008 Papers}, SIGGRAPH '08, pages 91:1--91:9,
  New York, NY, USA, 2008. ACM.

\bibitem{frohlich2011example}
S.~Fröhlich and M.~Botsch.
\newblock Example-driven deformations based on discrete shells.
\newblock {\em Computer Graphics Forum}, 30(8):2246--2257, 2011.

\bibitem{grinspun2003discrete}
E.~Grinspun, A.~N. Hirani, M.~Desbrun, and P.~Schr\"{o}der.
\newblock Discrete shells.
\newblock In {\em Proceedings of the 2003 ACM SIGGRAPH/Eurographics Symposium
  on Computer Animation}, SCA '03, pages 62--67, Aire-la-Ville, Switzerland,
  Switzerland, 2003. Eurographics Association.

\bibitem{jacobson2012fast}
A.~Jacobson, I.~Baran, L.~Kavan, J.~Popovi\'{c}, and O.~Sorkine.
\newblock Fast automatic skinning transformations.
\newblock {\em ACM Trans. Graph.}, 31(4):77:1--77:10, July 2012.

\bibitem{jacobson2011bounded}
A.~Jacobson, I.~Baran, J.~Popovi\'{c}, and O.~Sorkine-Hornung.
\newblock Bounded biharmonic weights for real-time deformation.
\newblock {\em Commun. ACM}, 57(4):99--106, Apr. 2014.

\bibitem{james2005skinning}
D.~L. James and C.~D. Twigg.
\newblock Skinning mesh animations.
\newblock In {\em ACM SIGGRAPH 2005 Papers}, SIGGRAPH '05, pages 399--407, New
  York, NY, USA, 2005. ACM.

\bibitem{kaufman1987clustering}
L.~Kaufman and P.~Rousseeuw.
\newblock {\em Clustering by means of medoids}.
\newblock North-Holland, 1987.

\bibitem{kavan2010fast}
L.~Kavan, P.-P. Sloan, and C.~O'Sullivan.
\newblock Fast and efficient skinning of animated meshes.
\newblock {\em Computer Graphics Forum}, 29(2):327--336, 2010.

\bibitem{kim2011blended}
V.~G. Kim, Y.~Lipman, and T.~Funkhouser.
\newblock Blended intrinsic maps.
\newblock In {\em ACM SIGGRAPH 2011 Papers}, SIGGRAPH '11, pages 79:1--79:12,
  New York, NY, USA, 2011. ACM.

\bibitem{koyama2012real}
Y.~Koyama, K.~Takayama, N.~Umetani, and T.~Igarashi.
\newblock Real-time example-based elastic deformation.
\newblock In {\em Proceedings of the 11th ACM SIGGRAPH / Eurographics
  Conference on Computer Animation}, EUROSCA'12, pages 19--24, Aire-la-Ville,
  Switzerland, Switzerland, 2012. Eurographics Association.

\bibitem{kry2002eigenskin}
P.~G. Kry, D.~L. James, and D.~K. Pai.
\newblock Eigenskin: Real time large deformation character skinning in
  hardware.
\newblock In {\em Proceedings of the 2002 ACM SIGGRAPH/Eurographics Symposium
  on Computer Animation}, SCA '02, pages 153--159, New York, NY, USA, 2002.
  ACM.

\bibitem{le2014robust}
B.~H. Le and Z.~Deng.
\newblock Robust and accurate skeletal rigging from mesh sequences.
\newblock {\em ACM Trans. Graph.}, 33(4):84:1--84:10, July 2014.

\bibitem{levi2015smooth}
Z.~Levi and C.~Gotsman.
\newblock Smooth rotation enhanced as-rigid-as-possible mesh animation.
\newblock {\em IEEE transactions on visualization and computer graphics},
  21(2):264--277, 2015.

\bibitem{lewis2000pose}
J.~P. Lewis, M.~Cordner, and N.~Fong.
\newblock Pose space deformation: A unified approach to shape interpolation and
  skeleton-driven deformation.
\newblock In {\em Proceedings of the 27th Annual Conference on Computer
  Graphics and Interactive Techniques}, SIGGRAPH '00, pages 165--172, New York,
  NY, USA, 2000. ACM Press/Addison-Wesley Publishing Co.

\bibitem{li2008global}
H.~Li, R.~W. Sumner, and M.~Pauly.
\newblock Global correspondence optimization for non-rigid registration of
  depth scans.
\newblock {\em Computer Graphics Forum}, 27(5):1421--1430, 2008.

\bibitem{lipman2004differential}
Y.~Lipman, O.~Sorkine, D.~Cohen-Or, D.~Levin, C.~Rossi, and H.~P. Seidel.
\newblock Differential coordinates for interactive mesh editing.
\newblock In {\em Shape Modeling Applications, 2004. Proceedings}, pages
  181--190, June 2004.

\bibitem{liu2008local}
L.~Liu, L.~Zhang, Y.~Xu, C.~Gotsman, and S.~J. Gortler.
\newblock A local/global approach to mesh parameterization.
\newblock {\em Computer Graphics Forum}, 27(5):1495--1504, 2008.

\bibitem{magnenat1988joint}
N.~Magnenat-thalmann, R.~Laperrire, D.~Thalmann, and U.~D. Montréal.
\newblock Joint-dependent local deformations for hand animation and object
  grasping.
\newblock In {\em In Proceedings on Graphics interface ’88}, pages 26--33,
  1988.

\bibitem{martin2011example}
S.~Martin, B.~Thomaszewski, E.~Grinspun, and M.~Gross.
\newblock Example-based elastic materials.
\newblock {\em ACM Trans. Graph.}, 30(4):72:1--72:8, July 2011.

\bibitem{mcadams2011computing}
A.~McAdams, A.~Selle, R.~Tamstorf, J.~Teran, and E.~Sifakis.
\newblock Computing the singular value decomposition of 3$\times$ 3 matrices
  with minimal branching and elementary floating point operations.
\newblock Technical report, Technical Report, University of Wisconsin-Madison,
  2011.

\bibitem{pinkall93}
U.~Pinkall and K.~Polthier.
\newblock Computing discrete minimal surfaces and their conjugates.
\newblock {\em Experimental Mathematics}, 2(1):15--36, 1993.

\bibitem{schumacher2012efficient}
C.~Schumacher, B.~Thomaszewski, S.~Coros, S.~Martin, R.~Sumner, and M.~Gross.
\newblock Efficient simulation of example-based materials.
\newblock In {\em Proceedings of the ACM SIGGRAPH/Eurographics Symposium on
  Computer Animation}, pages 1--8. Eurographics Association, 2012.

\bibitem{sloan2001shape}
P.-P.~J. Sloan, C.~F. Rose, III, and M.~F. Cohen.
\newblock Shape by example.
\newblock In {\em Proceedings of the 2001 Symposium on Interactive 3D
  Graphics}, I3D '01, pages 135--143, New York, NY, USA, 2001. ACM.

\bibitem{sorkine2007rigid}
O.~Sorkine and M.~Alexa.
\newblock As-rigid-as-possible surface modeling.
\newblock In {\em Proceedings of EUROGRAPHICS/ACM SIGGRAPH Symposium on
  Geometry Processing}, pages 109--116, 2007.

\bibitem{sumner2004deformation}
R.~W. Sumner and J.~Popovi\'{c}.
\newblock Deformation transfer for triangle meshes.
\newblock {\em ACM Trans. Graph.}, 23(3):399--405, Aug. 2004.

\bibitem{sumner2005mesh}
R.~W. Sumner, M.~Zwicker, C.~Gotsman, and J.~Popovi\'{c}.
\newblock Mesh-based inverse kinematics.
\newblock In {\em ACM SIGGRAPH 2005 Papers}, SIGGRAPH '05, pages 488--495, New
  York, NY, USA, 2005. ACM.

\bibitem{tycowicz2015interpolation}
C.~Von-Tycowicz, C.~Schulz, H.-P. Seidel, and K.~Hildebrandt.
\newblock Real-time nonlinear shape interpolation.
\newblock {\em ACM Trans. Graph.}, 34(3):34:1--34:10, May 2015.

\bibitem{wang2007real}
R.~Y. Wang, K.~Pulli, and J.~Popovi\'{c}.
\newblock Real-time enveloping with rotational regression.
\newblock {\em ACM Trans. Graph.}, 26(3), July 2007.

\bibitem{wang2015linear}
Y.~Wang, A.~Jacobson, J.~Barbi{\v{c}}, and L.~Kavan.
\newblock Linear subspace design for real-time shape deformation.
\newblock {\em ACM Transactions on Graphics (TOG)}, 34(4):57, 2015.

\bibitem{winkler2010multi}
T.~Winkler, J.~Drieseberg, M.~Alexa, and K.~Hormann.
\newblock Multi-scale geometry interpolation.
\newblock {\em Computer Graphics Forum}, 29(2):309--318, 2010.

\bibitem{xu2006poisson}
D.~Xu, H.~Zhang, Q.~Wang, and H.~Bao.
\newblock Poisson shape interpolation.
\newblock {\em Graphical Models}, 68(3):268--281, 2006.

\bibitem{zhang2008deformation}
H.~Zhang, A.~Sheffer, D.~Cohen-Or, Q.~Zhou, O.~Van~Kaick, and A.~Tagliasacchi.
\newblock Deformation-driven shape correspondence.
\newblock {\em Computer Graphics Forum}, 27(5):1431--1439, 2008.

\bibitem{zhang2015real}
W.~Zhang, J.~Zheng, and N.~M. Thalmann.
\newblock Real-time subspace integration for example-based elastic material.
\newblock In {\em Computer Graphics Forum}, volume~34, pages 395--404. Wiley
  Online Library, 2015.

\bibitem{zou2005regularization}
H.~Zou and T.~Hastie.
\newblock Regularization and variable selection via the elastic net.
\newblock {\em Journal of the Royal Statistical Society: Series B (Statistical
  Methodology)}, 67(2):301--320, 2005.

\end{thebibliography}
 }

\end{document}